

\input jytex
\typesize=10pt
\sectionnumstyle{arabic}
\footnoteskip=2pt
%
%

\def\PRL#1#2#3{{\sl Phys. Rev. Lett.} {\bf#1} (#2) #3}

\def\CMP#1#2#3{{\sl Comm. Math. Phys.} {\bf #1} (#2) #3}

\def\PRB#1#2#3{{\sl Phys. Rev.} {\bf B#1} (#2) #3}

\def\JMP#1#2#3{{\sl J. Math. Phys.} {\bf #1} (#2) #3}

\def\JPA#1#2#3{{\sl J. Physics} {\bf A#1} (#2) #3}
\def\JPC#1#2#3{{\sl J. Physics} {\bf C#1} (#2) #3}
\def\JSM#1#2#3{{\sl J. Soviet Math.} {\bf #1} (#2) #3}

\def\JETP#1#2#3{{\sl Sov. Phys. JETP} {\bf #1} (#2) #3}

\def\ZP#1#2#3{{\sl Z.Phys.} {\bf #1} (#2) #3}
\catcode`\@=11

\catcode`\^^?=13 \def^^?{\relax}


\newif\iftwoup
\ifx\@outputstyle\twoupoutput\twouptrue\fi

\iftwoup
     \topmargin=.5in
     \leftmargin=.3333in
     \vsize=7.25in
     \hsize=5in
     \fullhsize=10.3333in
\else\leftmargin=1.25in
     \vsize=9in
     \hsize=6in
\fi

\baselinestretch=1200
\parskip=0pt

\abovedisplayskip=10\rp@ plus3\rp@ minus5\rp@
\belowdisplayskip=10\rp@ plus3\rp@ minus5\rp@
\abovedisplayshortskip=5\rp@ plus2\rp@ minus4\rp@
\belowdisplayshortskip=10\rp@ plus3\rp@ minus5\rp@

\skip\footins=12\rp@ plus3\rp@ minus3\rp@

\hyphenation{pre-print}


\def\upref#1/{\markup{[\putref{#1}]}}

\newif\iftwosided
\def\twosided{\ifx\outputtype\twoupoutput\relax\else
     \twosidedtrue\evenleftmargin=1in \fi}



\jot=6\rp@

\everymath{\thinmuskip=3mu minus1mu \medmuskip=3mu plus1mu minus.5mu
     \thickmuskip=4mu plus2mu minus1mu }

\def~{\ifmmode\mskip6mu \relax\else\penalty10000 \ \fi}


\def\derLR{\rlap{\vbox to1em{%
     \hbox{\m@th$\scriptscriptstyle\leftrightarrow$}\vfil}}\partial}

\def\Idoubled#1{{\rm I\kern-.22em #1}}
\def\Odoubled#1{{\setbox0=\hbox{\rm#1}%
     \dimen@=\ht0 \dimen@ii=.04em \advance\dimen@ by-\dimen@ii
     \rlap{\kern.26em \vrule height\dimen@ depth-\dimen@ii width.075em}\box0}}

\def\Complex{\Odoubled C}


\catcode`\^^?=15

\catcode`\@=12

\def\sci#1#2#3{{\sl Science} {\bf #1} (#2) #3}
\def\vac{|0\rangle}
\def\vacn{|0\rangle^{(1)}}
\def\l#1{\lambda_{#1}}
\def\tl#1{{\tilde\lambda}_{#1}}
\def\la{\lambda}
\def\1l#1{\lambda^{(1)}_{#1}}
\def\2t#1{{\tilde\la}^{(1)}_{#1}}
\def\a#1#2{a({#1}-{#2})}
\def\b#1#2{b({#1}-{#2})}
\def\ta#1{{{#1}-{i\over 2}\over {#1}+{i\over 2}}}
\def\tb#1{{{#1}+{i\over 2}\over {#1}-{i\over 2}}}
\def\tc#1{{{#1}+{i}\over {#1}-{i}}}
\def\P{\Pi}
\def\Pg{\Pi}
\def\pg{\Pi^{(1)}}
\def\pbf{\Pi_{BF}}
\def\pfb{\Pi_{FB}}
\def\LE#1#2#3#4#5#6{{L^{(1)}_{#1}(#2 -\l{#1})}^{#3#4}_{#5#6}}
\def\Le1#1#2{L^{(1)}_{#1}(#2 -\l{#1})}
\def\ra#1#2#3#4#5#6{{r(#1-#2)^{#3#4}_{#5#6}\over \a#1#2 }}
\def\r#1#2#3#4#5#6{r(#1-#2)^{#3#4}_{#5#6}\ }
\def\rfb#1#2#3#4#5#6{r_{FB}(#1-#2)^{#3#4}_{#5#6}\ }

\def\R#1#2#3#4#5#6{R(#1-#2)^{#3#4}_{#5#6}\ }
\def\sr#1#2#3#4#5{r(#1)^{#2#3}_{#4#5}\ }
\def\srfb#1#2#3#4#5{r_{FB}(#1)^{#2#3}_{#4#5}\ }
\def\sR#1#2#3#4#5{R(#1)^{#2#3}_{#4#5}\ }
\def\ga{\alpha}
\def\gb{\beta}
\def\gc{\gamma}
\def\gd{\delta}
\def\ge{\epsilon}
\def\gs{\sigma}
\def\Sd{S^\dagger}
\def\Sz{S^z}
\def\qd{Q^\dagger}
\def\vmn{V^{(m|n)}}
\def\v12{V^{(1|2)}}
\def\osl{\bar\otimes}
\def\scr{\scriptstyle}
\def\da{\downarrow}

\equation{hintro}{H_{(3)} = i\sum_{k=1}^L \lbrack H_{susy}^{k,k+1},
H_{susy}^{k-1,k}\rbrack\quad ,}

\equation{config}{\vac \ , \qquad
|\uparrow\rangle _i = c^\dagger_{i,1} \vac \ , \qquad
|\downarrow\rangle _i = c^\dagger_{i,-1} \vac \quad .}

\equation{spin}{
S_j = c^\dagger_{j,1} c_{j,-1}\ , \qquad
S^\dagger_j = c^\dagger_{j,-1} c_{j,1}\ ,\qquad
S^z_j = {1\over 2} (n_{j,1}-n_{j,-1}) }

\equation{u12}{\eqalign{
J_{j,1} &= S_j = c^\dagger_{j,1} c_{j,-1}\cr
J_{j,2} &= S^\dagger_j =  c^\dagger_{j,-1} c_{j,1}\cr
J_{j,3} &= S^z_j = {1\over 2} (n_{j,1}-n_{j,-1})\cr
J_{j,4} &= Q_{j,1} = (1-n_{j,-1})\ c_{j,1}\cr
J_{j,5} &= Q^\dagger_{j,1} = (1-n_{j,-1})\  c^\dagger_{j,1}\cr}
\qquad\quad
\eqalign{
J_{j,6} &= Q_{j,-1} = (1-n_{j,1})\ c_{j,-1}\cr
J_{j,7} &= Q^\dagger_{j,-1} = (1-n_{j,1})\ c^\dagger_{j,-1} \cr
J_{j,8} &= T_j =1-{1\over 2}n_j\cr
J_{j,9} &= I_j=1\ \ .\cr}}

\equation{u12matrix}{
\eqalign{
J_{k,1} = S_k &= \left(\matrix{\scr 0&\scr 0&\scr 0\cr\scr 1&\scr 0&\scr
          0\cr \scr 0&\scr 0&\scr 0\cr}\right)\cr
J_{k,4} = Q_{k,1} &= \left(\matrix{\scr 0&\scr 0&\scr 0\cr \scr 0&\scr
          0&\scr 0\cr \scr 0&\scr 1&\scr 0\cr}\right)\cr
J_{k,7} = Q^\dagger_{k,-1} &= \left(\matrix{\scr 0&\scr 0&\scr 1\cr
          \scr 0&\scr 0&\scr 0\cr \scr 0&\scr 0&\scr 0\cr}\right)\cr}
\ \
\eqalign{
J_{k,2} = S_k^\dagger &= \left(\matrix{\scr 0&\scr 1&\scr 0\cr \scr 0&\scr
           0&\scr 0\cr \scr 0&\scr 0&\scr 0\cr}\right)\cr
J_{k,5} = Q^\dagger_{k,1} &= \left(\matrix{\scr 0&\scr 0&\scr 0\cr \scr
           0&\scr 0&\scr 1\cr\scr 0&\scr 0&\scr 0\cr}\right)\cr
J_{k,8} = T_k &= \left(\matrix{\scr {1\over 2}&\scr 0&\scr 0\cr \scr
          0&\scr {1\over 2}&\scr 0\cr \scr 0&\scr 0&\scr 1\cr}\right)\cr}
\ \
\eqalign{
J_{k,3} = S_k^z =& \left(\matrix{\scr -{1\over 2}&\scr 0&\scr  0\cr\scr
          0&\scr {1\over 2}&\scr 0\cr \scr 0&\scr 0&\scr 0\cr}\right)\cr
J_{k,6} = Q_{k,-1} &= \left(\matrix{\scr 0&\scr 0&\scr 0\cr\scr 0&\scr
          0&\scr 0\cr \scr 1&\scr 0&\scr 0\cr}\right)\cr
J_{k,9} = I_k &= \left(\matrix{\scr 1&\scr 0&\scr 0\cr \scr 0&\scr
          1&\scr 0\cr \scr 0&\scr 0&\scr 1\cr}\right)\cr}}

\equation{Kab}{K_{\ga\gb} = str(J_\ga\ J_\gb) = \left(\matrix{
\scr 0&\scr -1&\scr 0&\scr 0&\scr 0&\scr 0&\scr 0&\scr 0&\scr 0\cr
\scr -1&\scr 0&\scr 0&\scr 0&\scr 0&\scr 0&\scr 0&\scr 0&\scr 0\cr
\scr 0&\scr 0&\scr -{1\over 2}&\scr 0&\scr 0&\scr 0&\scr 0&\scr 0&\scr 0\cr
\scr 0&\scr 0&\scr 0&\scr 0&\scr 1&\scr 0&\scr 0&\scr 0&\scr 0\cr
\scr 0&\scr 0&\scr 0&\scr -1&\scr 0&\scr 0&\scr 0&\scr 0&\scr 0\cr
\scr 0&\scr 0&\scr 0&\scr 0&\scr 0&\scr 0&\scr 1&\scr 0&\scr 0\cr
\scr 0&\scr 0&\scr 0&\scr 0&\scr 0&\scr -1&\scr 0&\scr 0&\scr 0\cr
\scr 0&\scr 0&\scr 0&\scr 0&\scr 0&\scr 0&\scr 0&\scr {1\over 2}&\scr 0\cr
\scr 0&\scr 0&\scr 0&\scr 0&\scr 0&\scr 0&\scr 0&\scr 0&\scr
-1\cr}\right)\qquad .}

\equation{Hagain}{H_{susy}= \sum_{j=1}^L H_{susy}^{j,j+1}
= -\sum_{j=1}^L \Pg^{j,j+1}
= -\sum_{j=1}^L K^{\ga\gb}\ J_{j,\ga} \ J_{j+1,\gb}\qquad .}

\equation{H}{\eqalign{H_{susy}&= \sum_{j=1}^L H_{susy}^{j,j+1} = -\sum_{j=1}^L
K^{\ga\gb}\ J_{j,\ga} \ J_{j+1,\gb}\cr
&=\ -\sum_{j=1}^L\sum_{\sigma=\pm 1} Q^\dagger_{j,\sigma}
Q_{j+1,\sigma} + Q^\dagger_{j+1,\sigma} Q_{j,\sigma} \cr
&\qquad\qquad -\sum_{j=1}^L -2 S_j^z S_{j+1}^z - S^\dagger_j S_{j+1} - S_j
S^\dagger_{j+1} + 2 T_j T_{j+1} - I_j I_{j+1}\qquad ,\cr}}

\equation{h}{\eqalign{H&= -t\sum_{j=1}^L
\sum_{\sigma=\pm 1} c^\dagger_{j,\sigma}(1-n_{j,-\sigma})
c_{j+1,\sigma}(1-n_{j+1,-\sigma}) +
c^\dagger_{j+1,\sigma}(1-n_{j+1,-\sigma})
c_{j,\sigma}(1-n_{j,-\sigma}) \cr
&\qquad\qquad +J\sum_{j=1}^L
\left( S_j^z S_{j+1}^z +{1\over 2} \left(S^\dagger_j
S_{j+1} + S_j S^\dagger_{j+1}\right) -{1\over 4} n_j\
n_{j+1}\right)\qquad .\cr}}

\equation{hugly}{{H= \sum_{j=1}^L\left\lbrace
-t{\cal P} \sum_{\sigma=\pm 1} \left(c^\dagger_{j,\sigma} c_{j+1,\sigma} +
h.c.\right){\cal P} + J\left({\bf S}_j\cdot {\bf S}_{j+1}-{1\over 4} n_j\
n_{j+1}\right)\right\rbrace\quad ,}}

\equation{perm}{\eqalign{
\Pg^{j,j+1}\vac _j\times \vac _{j+1} &=
\vac _{j}\times \vac _{j+1}\cr
\Pg^{j,j+1}\vac _j\times |\sigma\rangle_{j+1} &=
|\sigma\rangle_{j}\times \vac _{j+1}\cr
\Pg^{j,j+1}|\tau\rangle_{j}\times |\sigma\rangle_{j+1} &=
- |\sigma\rangle_{j}\times |\tau\rangle_{j+1}
\quad ,\qquad \sigma ,\tau =\uparrow
,\downarrow\ .\cr}}

\equation{hpi}{\eqalign{H_{susy} &= H+2\hat N -L = -\sum_{j=1}^L
\Pg^{j,j+1}\cr &=\sum_{j=1}^L H_{susy}^{j,j+1}\qquad .\cr}}

\equation{gp}{\eqalign{\epsilon_1 &= \epsilon_1 = \epsilon_2 = \epsilon_3 =
\epsilon_8 = \epsilon_9 = 0 \cr
\epsilon_4 &= \epsilon_5 = \epsilon_6 = \epsilon_7 = 1 \qquad .\cr}}

\equation{kab}{K_{\ga\gb} = \left( K^{\ga\gb}\right)^{-1} = str(J_{j,\ga}\
J_{j,\gb})}

\equation{f}{ \lbrack J_{k,\ga}, J_{k,\gb}\rbrace := J_{k,\ga} J_{k,\gb} -
(-1)^{\epsilon_\ga\epsilon_\gb} J_{k,\gb} J_{k,\ga} =
{f_{\ga\gb}}^\gamma J_{k,\gamma}\qquad ,}

\equation{inv}{[H,J_\ga] = 0\qquad \ga=1,\ldots 9\qquad .}

\equation{linops}{M=\left(\matrix{A&B\cr C&D\cr}\right)\ ,\quad
\ge\left(\matrix{A&0\cr 0&D\cr}\right)=0\ ,\quad
\ge\left(\matrix{0&B\cr C&0\cr}\right)=1\quad .}

\equation{str}{str(M) = tr(A)-tr(D)\quad ,}

\equation{tensorprod}{v\otimes w = (e_av_a)\otimes (e_bw_b) =
(e_a\otimes e_b)v_aw_b (-1)^{\ge_{v_a}\ge_b}\quad .}

\equation{VV}{\left(R(\l{})\ e_j\otimes e_k\right)^{a_1}_{a_2} =
e_j^{b_1}\otimes e_k^{b_2}\ R(\l{})^{b_1a_1}_{b_2a_2}
\quad .}

\equation{linop}{\left(F\otimes G\right)(v\otimes w) = F(v)\otimes
G(w)\quad .}

\equation{linopmat}{\left(F\otimes G\right)^{ab}_{cd} = F_{ab}G_{cd}\
(-1)^{\ge_c(\ge_a+\ge_b)} \quad .}

\equation{pigr}{\Pg (v\otimes w) = (w\otimes v),\quad
(\Pg)^{a_1 b_1}_{a_2 b_2} = \delta_{a_1b_2}
\delta_{a_2 b_1} (-1)^{\epsilon_{b_1}\epsilon_{b_2}}\quad .}

\equation{AC}{\eqalign{A_{ab}(\mu)\ C_c(\l{}) &= (-1)^{\ge_a\ge_p}\
\ra{\mu}{\la}{d}{c}{p}{b}\ C_p(\l{})\ A_{ad}(\mu)\ +\ {\b{\mu}{\la}\over
\a{\mu}{\la}} C_{b}(\mu ) A_{ac}(\la )\cr
D(\mu)\ C_c(\l{}) &=
{1\over \a{\la}{\mu}}\ C_c(\l{})\ D(\mu)\ -\ {\b{\la}{\mu}\over
\a{\la}{\mu}} C_{c}(\mu ) D(\la )\cr
C_{a_1}(\l{1})\ C_{a_2}(\l{2}) &=
\r{\l{1}}{\l{2}}{b_1}{a_2}{b_2}{a_1}\ C_{b_2}(\l{2})\
C_{b_1}(\l{1})\quad ,\cr}}

\equation{r}{\eqalign{\sr{\mu}{a}{b}{c}{d} &= b({\mu})\delta_{ab}
\delta_{cd} - a({\mu})\delta_{ad}\delta_{bc}\cr
&= b({\mu})I^{ab}_{cd} + a({\mu}){\pg}^{ab}_{cd}\quad .\cr}}

\equation{ybe}{\r{\l{}}{\mu}{a_2}{c_2}{a_3}{c_3}
\sr{\l{}}{a_1}{b_1}{c_2}{d_2}\sr{\mu}{d_2}{b_2}{c_3}{b_3} =
\sr{\mu}{a_1}{c_1}{a_2}{c_2}\sr{\l{}}{c_2}{d_2}{a_3}{b_3}
\r{\l{}}{\mu}{c_1}{b_1}{d_2}{b_2}\qquad ,}

\equation{YBE}{\R{\l{}}{\mu}{a_2}{c_2}{a_3}{c_3}
\sR{\l{}}{a_1}{b_1}{c_2}{d_2}\sR{\mu}{d_2}{b_2}{c_3}{b_3} =
\sR{\mu}{a_1}{c_1}{a_2}{c_2}\sR{\l{}}{c_2}{d_2}{a_3}{b_3}
\R{\l{}}{\mu}{c_1}{b_1}{d_2}{b_2}\quad .}

\equation{YBE1}{\left(I\otimes R(\la -\mu)\right)
\left(R(\la )\otimes I\right)
\left(I\otimes R(\mu )\right) =
\left(R(\mu )\otimes I\right)
\left(I\otimes R(\la )\right)
\left(R(\la -\mu )\otimes I\right)
\quad .}

\equation{R}{R(\l{})=\left({\matrix{\scr
b(\l{})-a(\l{})&\scr 0&\scr 0&\scr
0&\scr 0& \scr 0&\scr 0&\scr
0&\scr 0\cr\scr 0&\scr b(\l{})&\scr
0&\scr -a(\l{})& \scr 0&\scr 0&\scr
0&\scr 0&\scr 0\cr\scr 0&\scr
0&\scr b(\l{})&\scr 0&\scr 0& \scr
0&\scr a(\l{})&\scr 0&\scr 0\cr\scr
0&\scr -a(\l{})&\scr
0&\scr b(\l{})& \scr 0&\scr 0&\scr
0&\scr 0&\scr 0\cr\scr 0&\scr
0&\scr 0&\scr 0&\scr
b(\l{})-a(\l{})&\scr 0&\scr 0&\scr
0&\scr 0\cr\scr 0&\scr 0&\scr
0&\scr 0&\scr 0&\scr
b(\l{})&\scr 0&\scr a(\l{})&\scr
0\cr\scr 0&\scr 0&\scr a(\l{})&\scr
0&\scr 0&\scr 0&\scr b(\l{})&\scr
0&\scr 0\cr\scr 0&\scr 0&\scr
0&\scr 0&\scr 0&\scr a(\l{})&\scr
0&\scr  b(\l{})&\scr 0\cr\scr  0&\scr
0&\scr 0&\scr 0&\scr 0&\scr
0&\scr 0&\scr 0&\scr
1\cr\scr}}\right)\ .}

\equation{YBE2}{R_{12}(\l{}-\mu)\left(\left\lbrack\P_{13}
R_{13}(\l{})\right\rbrack\otimes \left\lbrack\P_{23}
R_{23}(\mu)\right\rbrack\right) =
\left(\left\lbrack\P_{13} R_{13}(\mu)\right\rbrack\otimes
\left\lbrack\P_{23} R_{23}(\l{})\right\rbrack\right)
 R_{12}(\l{}-\mu)\quad ,}

\equation{LR}{L_n(\l{})^{ab}_{\alpha\beta}=\Pg^{ac}_{\ga\gc}
\sR{\l{}}{c}{b}{\gc}{\gb} = \left( b(\la )\Pg + a(\la )
I\right)^{ab}_{\ga\gb} \qquad .}

\equation{trid1}{H_{(2)} = -i\ {\partial ln(\tau (\l{}))\over
\partial\l{}}\bigg |_{\l{}=0} = -{\sum_{k=1}^L (\Pg^{k,k+1}-1)}.}

\equation{Rgen}{\eqalign{&R({\l{}}) = b(\l{})I + a(\l{}) \Pg\cr
&a(\l{}) = {\l{}\over \l{} +i}\ ,\qquad b(\l{}) = {i\over \l{} +i}\quad .}}

\equation{trid2}{H_{susy} = -i {\partial ln(\tau (\l{}))\over
\partial\l{}}\bigg |_{\l{}=0} - L\ =\ H_{(2)} -L\quad ,}

\equation{int}{\eqalign{&\R{\l{}}{\mu}{f_1}{c_1}{e_2}{c_2}\
\left(\Pg \ R(\l{})\right)^{c_1b_1}_{f_3 c_3}\
\left(\Pg \ R(\mu)\right)^{c_2 b_2}_{c_3 b_3}\
(-1)^{\ge_{c_2}(\ge_{c_1}+\ge_{b_1})} = \cr
&\left(\Pg \ R(\mu)\right)^{f_1 c_1}_{f_3 c_3}\
\left(\Pg \ R(\l{})\right)^{e_2c_2}_{c_3 b_3}\
\R{\l{}}{\mu}{c_1}{b_1}{c_2}{b_2}\
(-1)^{\ge_{e_2}(\ge_{f_1}+\ge_{c_1})}
\qquad .\cr}}

\equation{int1}{\eqalign{&\R{\l{}}{\mu}{a_1}{c_1}{a_2}{c_2}\
L_n(\l{})^{c_1b_1}_{\ga _n \gc_n}\
L_n(\mu)^{c_2 b_2}_{\gamma _n\beta _n}\
(-1)^{\ge_{c_2}(\ge_{c_1}+\ge_{b_1})}\ = \cr
&L_n(\mu)^{a_1 c_1}_{\ga _n \gc _n}\
L_n(\l{})^{a_2c_2}_{\gamma _n \beta _n}\
(-1)^{\ge_{a_2}(\ge_{a_1}+\ge_{c_1})}\
\R{\l{}}{\mu}{c_1}{b_1}{c_2}{b_2}\qquad .\cr}}

\equation{int2}{R(\l{}-\mu)\ \left(L_n(\l{})\otimes
L_n(\mu)\right) = \left(L_n(\mu)\otimes
L_n(\l{})\right)\ R(\l{}-\mu)\quad .}

\equation{intT}{R(\l{}-\mu)\ \left(T_L(\l{})\otimes
T_L(\mu)\right) = \left(T_L(\mu)\otimes
T_L(\l{})\right)\ R(\l{}-\mu)\quad .}

\equation{Lop}{L_n(\l{}) =
\left({\matrix{
a(\l{})-b(\l{})e_n^{11}&-b(\l{})e_n^{21}&b(\l{})e_n^{31}\cr
-b(\l{})e_n^{12}&a(\l{})-b(\l{})e_n^{22}&b(\l{})e_n^{32}\cr
b(\l{})e_n^{13}&b(\l{})e_n^{23}&a(\l{})+b(\la
)e_n^{33}\cr}}\right)\quad ,}

\equation{mon}{\eqalign{T_L(\l{}) &= L_L(\l{})L_{L-1}(\l{})\ldots
L_1(\l{})\cr
\left((T_L(\l{}))^{ab}\right)_{{\scriptstyle \ga_1\ldots\ga_L}\atop
{\scriptstyle \gb_1\ldots\gb_L}} &= L_L(\l{})^{ac_L}_{\ga_L\gb_L}
L_{L-1}(\l{})^{c_Lc_{L-1}}_{\ga_{L-1}\gb_{L-1}}\ldots
L_{1}(\l{})^{c_2b}_{\ga_{1}\gb_{1}}\quad\times\cr &\quad\times
(-1)^{\sum_{j=2}^L(\ge_{\ga_j}+\ge_{\gb_j})\sum_{i=1}^{j-1}
\ge_{\ga_i}}\quad .\cr}}

\equation{tau}{\tau (\l{}) = str(T_L(\l{})) = \sum_{a=1}^{m+n}
(-1)^{\ge_a} (T_L(\la ))^{aa}\qquad .}

\equation{T}{{T_L(\l{}) = L_L(\l{})L_{L-1}(\l{})\ldots
L_1(\l{})=\left(\matrix{A_{11}(\l{})&A_{12}(\l{})&B_1(\l{})\cr
A_{21}(\l{})&A_{22}(\l{})&B_2(\l{})\cr
C_1(\l{})&C_2(\l{})&D(\l{})\cr}\right)}\quad .}

\equation{vacuum}{{\vac _n = \pmatrix{0\cr 0\cr 1\cr}\qquad ,\qquad
\vac = \otimes_{n=1}^L \vac _n}\quad .}

\equation{Lvac}{L_k(\l{})\vac _k = \left(\matrix{a(\l{})&0&0\cr
0&a(\l{})&0\cr b(\l{})e_k^{13}&b(\l{})e_k^{23}&1}\right)\vac _k\qquad .}

\equation{Tvac}{T_L(\l{})\vac = \left(\matrix{(a(\l{}))^L&0&0\cr
0&(a(\l{}))^L&0\cr C_1(\l{})&C_2(\l{})&1}\right)\vac\qquad .}

\equation{state1}{|\l{1},\ldots , \l{n}|F\rangle = C_{a_1}(\l{1})\
C_{a_2}(\l{2})\ldots C_{a_n}(\l{n})\ \vac\  F^{a_n\ldots
a_1}\qquad ,}

\equation{ew}{\tau (\mu)|\l{1},\ldots , \l{n}|F\rangle = \nu
(\mu,\lbrace\l{j}\rbrace,F)\ |\l{1},\ldots , \l{n}|F\rangle }

\equation{tau2}{\tau (\mu) = str(T_L(\mu )) = -A_{11}(\mu )-A_{22}(\mu
)+ D(\mu )\qquad .}

\equation{Dstate}{\eqalign{D(\mu )|\l{1},\ldots , \l{n}|F\rangle =
&\prod_{j=1}^n {1\over \a{\l{j}}{\mu}} |\l{1},\ldots , \l{n}|F\rangle\cr
&+\sum_{k=1}^n\left({\tilde\Lambda}_k\right)^{b_1\ldots b_n}_{a_1\ldots a_n}
C_{b_k}(\mu )\prod_{\scr j=1\atop\scr j\neq k}^n C_{b_j}(\l{j})\vac
F^{a_n\ldots a_1}\quad ,\cr}}

\equation{Astate}{\eqalign{
(A_{11}(\mu )+A_{22}(\mu ))|\l{1},&\ldots ,
\l{n}|F\rangle =\cr
&=-\ (a(\mu ))^L\prod_{j=1}^n {1\over \a{\mu}{\l{j}}}
\prod_{l=1}^n C_{b_l}(\l{l})\vac\ \tau^{(1)}(\mu )^{b_1\ldots
b_n}_{a_1\ldots a_n}\ F^{a_n\ldots a_1} \cr
&\quad +\sum_{k=1}^n \left(\Lambda_k\right)^{b_1\ldots b_n}_{a_1\ldots a_n}
C_{b_k}(\mu )
\prod_{\scr j=1\atop\scr j\neq k}^n
C_{b_j}(\l{j})\vac F^{a_n\ldots a_1}\quad ,\cr}}

\equation{tau1}{\eqalign{\tau^{(1)}(\mu )^{b_1\ldots
b_n}_{a_1\ldots a_n} &= str( T_n^{(1)}(\mu )) \cr
&= str(\Le1{n}{\mu}\Le1{n-1}{\mu}\ldots\Le1{2}{\mu}\Le1{1}{\mu})\quad
,\cr
&{\hskip -80pt\rm and}\cr
L_k^{(1)}(\la) &= b(\la) \pg + a(\la) I^{(1)}\cr
&= \pg\ r(\la) = \left(\matrix{a(\la)-b(\la)\ e_k^{11}&-b(\la)\
e_k^{21}\cr -b(\la)\ e_k^{12}&a(\la)-b(\la)e_k^{22}\cr}\right)\quad
.\cr}}

\equation{unw}{\left(-(\Lambda_k)^{b_1\ldots b_n}_{a_1\ldots a_n} +
(\tilde\Lambda_k)^{b_1\ldots b_n}_{a_1\ldots a_n}\right)
F^{a_n\ldots a_1} = 0\quad .}

\equation{PBC}{(a(\l{k}))^{-L} \prod_{\scr
l=1\atop\scr l\neq k}^n
{\a{\l{k}}{\l{l}}\over\a{\l{l}}{\l{k}}}\ F^{b_n\ldots b_1} =
\tau^{(1)}(\l{k})^{b_1\ldots b_n}_{a_1\ldots a_n}\ F^{a_n\ldots
a_1}\quad ,\quad k=1,\ldots ,n\quad .}


\equation{Tnest}{T_n^{(1)}(\la) = \Le{n}{\la}\Le{n-1}{\la}\ldots
\Le{2}{\la}\Le{1}{\la}}

\equation{intnest}{r(\l{}-\mu)\left(T_n^{(1)}(\l{})\otimes
T_n^{(1)}(\mu)\right)=\left(T_n^{(1)}(\mu)\otimes
T_n^{(1)}(\l{})\right) r(\l{}-\mu)\quad .}

\equation{T1}{T^{(1)}_n(\mu ) = \left(\matrix{A^{(1)}(\mu)&B^{(1)}(\mu
)\cr
C^{(1)}(\mu )&D^{(1)}(\mu )\cr}\right)\quad ,\tau^{(1)}(\mu ) =
-A^{(1)}(\mu ) - D^{(1)}(\mu )\quad ,}

\equation{intnest2}{\eqalign{D^{(1)}(\mu ) C^{(1)}(\la ) &= {1\over
\a{\mu}{\la} } C^{(1)}(\la )D^{(1)}(\mu ) + {\b{\la}{\mu}\over
\a{\la}{\mu}} C^{(1)}(\mu )D^{(1)}(\la )\cr
A^{(1)}(\mu ) C^{(1)}(\la ) &= {1\over
\a{\la}{\mu} } C^{(1)}(\la )A^{(1)}(\mu ) + {\b{\mu}{\la}\over
\a{\mu}{\la}} C^{(1)}(\mu )A^{(1)}(\la )\cr
C^{(1)}(\la)C^{(1)}(\mu) &= C^{(1)}(\mu)C^{(1)}(\la)\quad .\cr}}

\equation{vacnest}{{\vac ^{(1)}_k = \pmatrix{0\cr 1\cr}\qquad ,\qquad
\vac ^{(1)} = \otimes_{k=1}^n \vac ^{(1)}_k}\quad .}

\equation{tvac}{\eqalign{A^{(1)}(\mu)\vac ^{(1)} &= \prod_{j=1}^n
\a{\mu}{\l{j}}\vacn\cr
D^{(1)}(\mu)\vac ^{(1)} &= \prod_{j=1}^n
\left(\a{\mu}{\l{j}}-\b{\mu}{\l{j}}\right)\vacn = \prod_{j=1}^n
{\a{\mu}{\l{j}}\over \a{\l{j}}{\mu}}\vacn\quad .\cr}}

\equation{state2}{|\1l{1},\ldots , \1l{n_1}\rangle =
C^{(1)}(\1l{1})\ C^{(1)}(\1l{2})\ldots C^{(1)}(\1l{n_1})
\ \vac\qquad .}

\equation{ADstate1}{\eqalign{
D^{(1)}(\mu )|\1l{1},\ldots , \1l{n_1}\rangle &=
\prod_{j=1}^{n_1} {1\over \a{\mu}{\1l{j}}}
\prod_{l=1}^n {\a{\mu}{\l{l}}\over \a{\l{l}}{\mu}}
|\1l{1},\ldots , \1l{n_1}\rangle\cr
&+\sum_{k=1}^{n_1} {\tilde\Lambda}_k^{(1)}\
C^{(1)}(\mu )\prod_{\scr j=1\atop\scr j\neq
k}^{n_1} C^{(1)}(\1l{j})\vacn\quad , \cr}}

\equation{ADstate2}{\eqalign{
A^{(1)}(\mu )|\1l{1},\ldots , \1l{n_1}\rangle &=
\prod_{j=1}^{n_1} {1\over \a{\1l{j}}{\mu}}
\prod_{l=1}^n \a{\mu}{\l{l}}
|\1l{1},\ldots , \1l{n_1}\rangle\cr
&+\sum_{k=1}^{n_1} {\Lambda_k^{(1)}}\
C^{(1)}(\mu )\prod_{\scr j=1\atop\scr j\neq
k}^{n_1} C^{(1)}(\1l{j})\vacn \quad .\cr}}

\equation{pbc}{\prod_{i=1}^n \a{\l{i}}{\1l{p}} = \prod_{\scr
j=1\atop\scr j\neq p}^{n_1}
{\a{\1l{j}}{\1l{p}}\over\a{\1l{p}}{\1l{j}}}\quad  ,\qquad p=1,\ldots
,n_1\ .}

\equation{eig}{\eqalign{
&\tau^{(1)} (\mu ) |\1l{1},\ldots , \1l{n_1}\rangle = \cr
&-\left(\prod_{i=1}^{n_1} {1\over \a{\mu}{\1l{i}}}\prod_{j=1}^n
{\a{\mu}{\l{j}}\over \a{\l{j}}{\mu}}
+\prod_{i=1}^{n_1} {1\over \a{\1l{i}}{\mu}}
\prod_{j=1}^n \a{\mu}{\l{j}}\right) |\1l{1},\ldots , \1l{n_1}\rangle
\quad .}}

\equation{PBC2}{\left(a(\l{k})\right)^L =
\prod_{i=1}^{n_1}\a{\l{k}}{\1l{i}}\quad ,\qquad k=1,\ldots
,n\ .}

\equation{BE}{\eqalign{
\left({\tl{k}-{i\over 2}\over\tl{k}+{i\over 2}}\right)^L &=
\prod_{j=1}^{N_{\downarrow}}{\tl{k} -\1l{j} - {i\over 2}
\over\tl{k} -\1l{j} + {i\over 2}}\quad ,\qquad k=1,\ldots ,{N_e}\cr
\prod_{k=1}^{N_e}{\tl{k} -\1l{p} - {i\over 2}
\over\tl{k} -\1l{p} + {i\over 2}} &=
\prod_{\scr j=1\atop\scr j\neq p}^{N_{\downarrow}}
{\1l{j} -\1l{p} -i\over\1l{j} - \1l{p} +i}
\quad ,\qquad p=1,\ldots ,N_{\downarrow}\ .\cr}}

\equation{evtau}{\eqalign{\nu (\mu,\lbrace\l{j}\rbrace ,F) &=
(a(\mu ))^L\prod_{j=1}^{N_e} {1\over \a{\mu}{\l{j}}} \nu ^{(1)}(\mu )\
+ \ \prod_{j=1}^{N_e} {1\over \a{\l{j}}{\mu}}\cr
\nu ^{(1)}(\mu ) &=
-\left(\prod_{i=1}^{N_{\downarrow}} {1\over
\a{\mu}{\1l{i}}}\prod_{j=1}^{N_e}
{\a{\mu}{\l{j}}\over \a{\l{j}}{\mu}}
+\prod_{i=1}^{N_{\downarrow}} {1\over \a{\1l{i}}{\mu}}
\prod_{j=1}^{N_e} \a{\mu}{\l{j}}\right) .\cr}}

\equation{energy}{E_{susy} = \sum_{j=1}^{N_e} {1\over {\tilde \l{j}}^2
+{1\over 4}} -L\ = \ -2\sum_{j=1}^{N_e}cos(k_j)\ + 2N_e-L\quad ,}


\equation{invr}{\r{\l{1}}{\l{2}}{b_1}{a_1}{b_2}{a_2}\
\r{\l{2}}{\l{1}}{c_1}{b_1}{c_2}{b_2} = I^{a_1 c_1}_{a_2 c_2} = \gd
_{a_1 c_1}\ \gd _{a_2 c_2}\quad ,}

\equation{front}{\eqalign{\prod_{i=1}^n C_{a_i} (\l{i}) &= C_{b_k} (\l{k})\
\prod_{i=1}^{k-1} C_{b_i}(\l{i})\ \prod_{j=k+1}^n C_{a_j}
(\l{j})\ S(\l{k})^{b_1\ldots b_k}_{a_1\ldots a_k}\quad
,\cr\noalign{\vskip 4pt}
S(\l{k})^{b_1\ldots b_k}_{a_1\ldots
a_k}&=\r{\l{k-1}}{\l{k}}{b_{k-1}}{a_k}{c_{k-1}}{a_{k-1}}\
\r{\l{k-2}}{\l{k}}{b_{k-2}}{c_{k-1}}{c_{k-2}}{a_{k-2}}\ldots
\r{\l{1}}{\l{k}}{b_{1}}{c_2}{b_{k}}{a_{1}}\quad .\cr}}

\equation{laktilde}{\left({\tilde\Lambda}_k F\right)^{b_1\ldots b_n}
= S(\l{k})^{b_1\ldots b_k}_{a_1\ldots a_k}\ F^{b_n\ldots
b_{k+1}a_k\ldots a_1}
\left(-{\b{\l{k}}{\mu}\over\a{\l{k}}{\mu}}\right)\prod_{\scr
i=1\atop\scr i\neq k}^n {1\over \a{\l{i}}{\l{k}}}
\quad .}

\equation{lak1}{\eqalign{
\left({\Lambda}_{k,1} F\right)^{b_1\ldots b_n}
&= S(\l{k})^{c_1\ldots c_k}_{a_1\ldots a_k}\ F^{a_n\ldots a_1}
\left({\b{\mu}{\l{k}}\over\a{\mu}{\l{k}}}\right) \gd_{b_k,1}\
\prod_{\scr
i=1\atop\scr i\neq k}^n {1\over \a{\l{k}}{\l{i}}}
\quad \times\cr
&\times \
\r{\l{k}}{\l{1}}{d_1}{c_1}{b_1}{c_k}
\r{\l{k}}{\l{2}}{d_2}{c_2}{b_2}{d_1}\ldots
\r{\l{k}}{\l{k-1}}{d_{k-1}}{c_{k-1}}{b_{k-1}}{d_{k-2}}\quad\times\cr
&\times \
\r{\l{k}}{\l{k+1}}{d_k}{a_{k+1}}{b_{k+1}}{d_{k-1}}
\r{\l{k}}{\l{k+2}}{d_{k+1}}{a_{k+2}}{b_{k+2}}{d_{k}}\ldots
\r{\l{k}}{\l{n}}{d_{n-1}}{a_{n}}{b_{n}}{d_{n-2}}\quad\times\cr
&\times (a(\l{k}))^L \ \gd_{d_{n-1},1}\ (-1)^{n-1}\quad .\cr}}

\equation{lak2}{\eqalign{
\left(\Lambda_k F\right)^{b_1\ldots b_n}
&= S(\l{k})^{c_1\ldots c_k}_{a_1\ldots a_k}\ F^{a_n\ldots a_1}
\left({\b{\mu}{\l{k}}\over\a{\mu}{\l{k}}}\right)\prod_{\scr
i=1\atop\scr i\neq k}^n {1\over \a{\l{k}}{\l{i}}}
\ (a(\l{k}))^L \quad \times\cr
&\times \
\r{\l{k}}{\l{1}}{d_1}{c_1}{b_1}{c_k}
\r{\l{k}}{\l{2}}{d_2}{c_2}{b_2}{d_1}\ldots
\r{\l{k}}{\l{k-1}}{d_{k-1}}{c_{k-1}}{b_{k-1}}{d_{k-2}}\ (-1)^{n-1}
\quad\times\cr
&\times \
\r{\l{k}}{\l{k+1}}{d_k}{a_{k+1}}{b_{k+1}}{d_{k-1}}
\r{\l{k}}{\l{k+2}}{d_{k+1}}{a_{k+2}}{b_{k+2}}{d_{k}}\ldots
\r{\l{k}}{\l{n}}{b_{k}}{a_{n}}{b_{n}}{d_{n-2}}\quad .\cr}}

\equation{raise}{r(\la )^{ab}_{cd} = (r(\la )\pg)^{ad}_{cb}\
(-1)^{\ge_a\ge_c} = -(r(\la )\pg)^{ad}_{cb} =
-L^{(1)}(\la )^{ad}_{cb}\quad .}

\equation{lak3}{\eqalign{
&\left({\Lambda}_k F\right)^{b_1\ldots b_n}
= {\b{\mu}{\l{k}}\over\a{\mu}{\l{k}}}\prod_{\scr
i=1\atop\scr i\neq k}^n {1\over \a{\l{k}}{\l{i}}}
\ (a(\l{k}))^L \ F^{a_n\ldots a_kb_{k-1}\ldots b_1}\ (-1)^{k+1}
\quad\times\cr
&\times\
\LE{{n}}{\l{k}}{b_{k}}{d_{n-2}}{b_{n}}{a_{n}}
\LE{{n-1}}{\l{k}}{d_{n-2}}{d_{n-3}}{b_{n-1}}{a_{n-1}}\ldots
\LE{{k+1}}{\l{k}}{d_k}{a_{k}}{b_{k+1}}{a_{k+1}}
\ \ .\cr}}

\equation{anti}{{\a{\mu}{\l{k}}-\b{\mu}{\l{k}}} =
{\a{\mu}{\l{k}}\over\a{\l{k}}{\mu}}\quad .}

\equation{cancel}{\eqalign{&\prod_{\scr
i=1\atop\scr i\neq k}^n {\a{\l{k}}{\l{i}}\over \a{\l{i}}{\l{k}}}
\ (a(\l{k}))^{-L} \ F^{b_n\ldots b_{k+1}p_{k}\ldots p_1}=  \cr
&=\LE{n}{\l{k}}{b_{k}}{d_{n-2}}{b_{n}}{a_{n}}\ldots
\LE{k+1}{\l{k}}{d_k}{a_{k}}{b_{k+1}}{a_{k+1}}\quad\times\cr
&\qquad\times\quad\r{\l{k}}{\l{k-1}}{p_k}{b_{k-1}}{p_{k-1}}{s_{k-1}}\ldots
\r{\l{k}}{\l{1}}{s_2}{b_{1}}{p_{1}}{b_{k}}\ F^{a_n\ldots
a_kb_{k-1}\ldots b_1}\ (-1)^{k+1}\cr
&=\LE{k-1}{\l{k}}{p_k}{s_{k-1}}{p_{k-1}}{b_{k-1}}
\LE{k-2}{\l{k}}{s_{k-1}}{s_{k-2}}{p_{k-2}}{b_{k-2}}\ldots
\LE{1}{\l{k}}{s_2}{b_{k}}{p_{1}}{b_{1}}\quad\times\cr
&\qquad\times\quad
\LE{n}{\l{k}}{b_k}{d_{n-2}}{b_{n}}{a_{n}}\ldots
\LE{k+1}{\l{k}}{d_k}{a_{k}}{b_{k+1}}{a_{k+1}}\
F^{a_n\ldots a_kb_{k-1}\ldots b_1}\cr
&=\left(\tau^{(1)}(\l{k})\ F\right)^{b_n\ldots b_{k+1}p_k\ldots p_1}
\quad .\cr}}

\equation{Sr}{S(\l{k})^{c_1\ldots c_k}_{a_1\ldots a_k}\
\r{\l{k}}{\l{1}}{d_1}{c_1}{b_1}{c_k}\ldots
\r{\l{k}}{\l{k-1}}{d_{k-1}}{c_{k-1}}{b_{k-1}}{d_{k-2}} =
\prod_{i=1}^{k-1}\gd_{a_i,b_i}\ \gd_{d_{k-1},a_k}\quad.}


\equation{1lak}{\eqalign{
\Lambda^{(1)}_k &= {\b{\mu}{\1l{k}}\over
\a{\mu}{\1l{k}}} \prod_{\scr j=1\atop\scr j\neq
k}^{n_1} {1\over \a{\1l{j}}{\1l{k}}}\prod_{l=1}^{n}\a{\1l{k}}{\l{l}}\cr
{\tilde\Lambda}^{(1)}_k &= {\b{\1l{k}}{\mu}\over
\a{\1l{k}}{\mu}} \prod_{\scr j=1\atop\scr j\neq
k}^{n_1} {1\over
\a{\1l{k}}{\1l{j}}}\prod_{l=1}^{n}{\a{\1l{k}}{\l{l}}\over
\a{\l{l}}{\1l{k}}}\quad .\cr}}


\equation{hcl}{ln\left(\tau(\l{})(\tau (0))^{-1}\right) =
\sum_{k=1}^\infty i{\l{}^k\over k!}
H_{(k+1)}\qquad .}

\equation{boost}{B = \sum_n n H_{(2)}^{n,n+1}\qquad ,}

\equation{hcl2}{\eqalign{H_{(k+1)} &= i\lbrack B,H_{(k)}\rbrack \cr
& =i\lbrack {\tilde B},H_{(k)}\rbrack\qquad ,\cr}}

\equation{boost2}{{\tilde B} = \sum_n n H_{susy}^{n,n+1}\qquad .}

\equation{Rtilde}{{\tilde R}(\la ) = \Pg R(\la ) = b(\la )\Pg + a(\la
)I\quad ,}

\equation{rllkul}{{\tilde R}(\l{}-\mu)(L_n(\l{})\otimes
L_n(\mu)) = \left(I\otimes L_n(\mu)\right)\left(L_n(\l{})\otimes
I)\right){\tilde R}(\l{}-\mu)\quad .}

\equation{90rot}{\eqalign{{\tilde R}_{n,n+1}(\la -\mu )&\left(
L_{n,1}(\la )\otimes L_{n+1,1}(\mu )\right) =\cr
&\left(I_n\otimes L_{n+1,1}(\mu )\right)\left(L_{n,1}(\la )\otimes
I_{n+1}\right) {\tilde R}_{n,n+1}(\la -\mu )\quad ,\cr}}

\equation{90rot1}{\eqalign{&{\tilde R}_{n,n+1}(\la
-\mu)^{\ga_1\gc_1}_{\ga_2\gc_2}\ L_{n}(\la )^{a_1b_1}_{\gc_1\gb_1}\
L_{n+1}(\mu )^{b_1c_1}_{\gc_2\gb_2}\
(-1)^{\ge_{\gc_2}(\ge_{\gc_1}+\ge_{\gb_1})} = \cr
&L_{n+1}(\mu )^{a_1b_1}_{\ga_2\gc_2}\
L_{n}(\la )^{b_1c_1}_{\ga_1\gc_1}\
(-1)^{\ge_{\gc_2}(\ge_{\ga_1}+\ge_{\gc_1})}
{\tilde R}_{n,n+1}(\la -\mu )^{\gc_1\gb_1}_{\gc_2\gb_2}
\quad .\cr}}

\equation{90rot2}{{\tilde R}_{n,n+1}(\la -\mu )\ L_{n}(\la )\ L_{n+1}(\mu )
= L_{n+1}(\mu )\ L_{n}(\la )\ {\tilde R}_{n,n+1}(\la -\mu )\quad .}

\equation{diff}{\eqalign{
{\partial\over\partial\la}\left((\la +i) {\tilde R}_{n,n+1}(\la
)\right) &= I^{n,n+1}\cr
{\tilde R}_{n,n+1}(0) &= \Pg^{n,n+1}\quad .\cr}}

\equation{suth}{\left\lbrack \Pg^{n,n+1}, L_{n+1}(\mu )\otimes
L_{n}(\mu ) \right\rbrack = -i{\dot L}_{n+1}(\mu )\otimes L_{n}(\mu )
+i L_{n+1}(\mu )\otimes {\dot L}_{n}(\mu )\quad .}

\equation{suth2}{\left\lbrack H_{(2)}^{n,n+1}, L_{n+1}(\mu )\otimes
L_{n}(\mu ) \right\rbrack = i\left({\dot L}_{n+1}(\mu )\otimes
L_{n}(\mu ) - L_{n+1}(\mu )\otimes {\dot L}_{n}(\mu )\right)\quad .}

\equation{hcl-1}{\left\lbrack B,\tau (\mu )\right\rbrack = -i\ {\dot
\tau}(\mu )}

\equation{hcl0}{\left\lbrack B,ln(\tau \left(\mu )(\tau
(0))^{-1}\right)\right\rbrack =
-i{\partial\over\partial\mu} ln\left(\tau (\mu )\ (\tau
(0))^{-1}\right) - H_{(2)}\quad .}

\equation{H3}{\eqalign{H_{(3)} &= i\ \lbrack \tilde B, H_{(2)}\rbrack =
i\ \lbrack \tilde B, H_{susy}\rbrack \cr
&=i\ \sum_{k=1}^L  \lbrack H_{susy}^{k+1,k+2},
H_{susy}^{k,k+1}\rbrack\cr
&=-i\ \sum_{k=1}^L K^{\ga\gb}\ K^{\gamma\gd}\ {f_{\gb\gamma}}^\ge
J_{k-1,\ga}\ J_{k,\ge}\ J_{k+1,\gd}\qquad .\cr}}

\equation{H3again}{\eqalign{H_{(3)}
=i\sum_{j=1}^L\sum_{k=1}^L
&\ -2 \Sd_{j-1}\ S_j\ \Sz_{j+1}
-2 \Sd_{j-1}\ \Sz_j\ S_{j+1}\cr\noalign{\vskip -8pt}
&\ + \Sd_{j-1}\ \qd_{j,1}\ Q_{j+1,-1}
+ \Sd_{j-1}\ Q_{j,-1}\ \qd_{j+1,1}\cr
&\ -2 \Sz_{j-1}\ \Sd_j\ S_{j+1}
+ (1-n_{j,-1})\ Q_{j,-1}\ \qd_{j+1,-1}
+ (1-n_{j,1})\ \qd_{j,1}\ Q_{j+1,1}\cr
&\ -\qd_{j-1,1}\ \Sd_j\ Q_{j+1,-1}
-\qd_{j-1,1}\ (1-n_{j,-1})\ Q_{j+1,1}\cr
&\ +\qd_{j-1,1}\ Q_{j,1}\ (1-n_{j,1})
+\qd_{j-1,1}\ Q_{j,-1}\ \Sd_{j+1}\cr
&\ -\qd_{j-1,-1}\ S_j\ Q_{j+1,1}
-\qd_{j-1,-1}\ (1-n_{j,1})\ Q_{j+1,-1}\cr
&\ +\qd_{j-1,-1}\ Q_{j,-1}\ (1-n_{j,-1})
+\qd_{j-1,-1}\ Q_{j,1}\ S_{j+1}\cr
&\ - h.c.\quad .\cr}}

\equation{H4}{\eqalign{H_{(4)} &= i\lbrack \tilde B, H_{(3)}\rbrack
\cr
&=-2\sum_{k=1}^L K^{\mu\nu}\ K^{\ga\gb}\ K^{\gamma\gd}\
{f_{\gb\gamma}}^\ge\ {f_{\gd\mu}}^\omega J_{k-1,\ga}\ J_{k,\ge}\
J_{k+1,\omega}\ J_{k+2,\nu}\cr
&\qquad +\sum_{k=1}^L { P}^{k-1,k+1}\ -2\sum_{k=1}^L \Pg^{k,k+1}
\qquad ,\cr}}

\equation{P}{{P}^{k-1,k+1} =
\Pg^{k-1,k}\Pg^{k,k+1}\Pg^{k-1,k}\quad .}

\equation{h3}{H_{(3)} = \sum_{n}\sum_{k=1}^L n \lbrack H_{susy}^{n,n+1},
H_{susy}^{k,k+1}\rbrack\quad .}


\equation{vacuum2}{{\vac _n = \pmatrix{0\cr 0\cr 1\cr}\qquad ,\qquad
\vac = \otimes_{n=1}^L \vac _n}}

\equation{R2}{R(\l{})=\left({\matrix{
\scr 1&\scr 0&\scr 0&\scr 0&\scr 0& \scr 0&\scr 0&\scr 0&\scr 0\cr
\scr 0&\scr b(\l{})&\scr 0&\scr a(\l{})& \scr 0&\scr 0&\scr 0&\scr
     0&\scr 0\cr
\scr 0&\scr 0&\scr b(\l{})&\scr 0&\scr 0& \scr 0&\scr a(\l{})&\scr
     0&\scr 0\cr
\scr 0&\scr a(\l{})&\scr 0&\scr b(\l{})& \scr 0&\scr 0&\scr 0&\scr
     0&\scr 0\cr
\scr 0&\scr 0&\scr 0&\scr 0&\scr b(\l{})-a(\l{})&\scr 0&\scr 0&\scr
     0&\scr 0\cr
\scr 0&\scr 0&\scr 0&\scr 0&\scr 0&\scr b(\l{})&\scr 0&\scr
     -a(\l{})&\scr 0\cr
\scr 0&\scr 0&\scr a(\l{})&\scr 0&\scr 0&\scr 0&\scr b(\l{})&\scr
     0&\scr 0\cr
\scr 0&\scr 0&\scr 0&\scr 0&\scr 0&\scr -a(\l{})&\scr 0&\scr
     b(\l{})&\scr 0\cr
\scr 0&\scr 0&\scr 0&\scr 0&\scr 0&\scr 0&\scr 0&\scr 0&\scr
     b(\la )-a(\la )\cr }}\right)\ .}

\equation{Lop2}{L_n(\l{}) =
\left({\matrix{
a(\l{})+b(\l{})e_n^{11}&b(\l{})e_n^{21}&b(\l{})e_n^{31}\cr
b(\l{})e_n^{12}&a(\l{})-b(\l{})e_n^{22}&-b(\l{})e_n^{32}\cr
b(\l{})e_n^{13}&-b(\l{})e_n^{23}&a(\l{})-b(\la )e_n^{33}\cr}}
\right)\quad .}

\equation{T2}{{T_L(\l{}) = L_L(\l{})L_{L-1}(\l{})\ldots
L_1(\l{})=\left(\matrix{A_{11}(\l{})&A_{12}(\l{})&B_1(\l{})\cr
A_{21}(\l{})&A_{22}(\l{})&B_2(\l{})\cr
C_1(\l{})&C_2(\l{})&D(\l{})\cr}\right)}\quad ,}

\equation{tau22}{\tau (\mu) = A_{11}(\mu )-A_{22}(\mu )- D(\mu )\qquad
.}

\equation{Lvac2}{L_k(\l{})\vac _k = \left(\matrix{a(\l{})&0&0\cr
0&a(\l{})&0\cr b(\l{})e_k^{13}&-b(\l{})e_k^{23}&a(\la )-b(\la )
}\right)\vac _k\qquad .}

\equation{Tvac2}{T_L(\l{})\vac = \left(\matrix{(a(\l{}))^L&0&0\cr
0&(a(\l{}))^L&0\cr C_1(\l{})&C_2(\l{})&\left(a(\la )-b(\la
)\right)^L }\right)\vac\qquad ,}

\equation{state12}{|\l{1},\ldots , \l{n}|F\rangle = C_{a_1}(\l{1})\
C_{a_2}(\l{2})\ldots C_{a_n}(\l{n})\ \vac \ F^{a_n\ldots
a_1}\qquad .}

\equation{AC2}{\eqalign{A_{ab}(\mu)\ C_c(\l{}) &=
(-1)^{\ge_a\ge_p+\ge_a+\ge_b}\ \ra{\mu}{\la}{d}{c}{p}{b}\ C_p(\l{})\
A_{ad}(\mu)\cr
&+ (-1)^{(\ge_a+1)(\ge_b+1)}\ {\b{\mu}{\la}\over
\a{\mu}{\la}} C_{b}(\mu ) A_{ac}(\la )\cr
D(\mu)\ C_c(\l{}) &=
{1\over \a{\mu}{\la}}\ C_c(\l{})\ D(\mu)\ +\ {\b{\la}{\mu}\over
\a{\la}{\mu}} C_{c}(\mu ) D(\la )\cr
C_{a_1}(\l{1})\ C_{a_2}(\l{2}) &=
\rfb{\l{2}}{\l{1}}{a_2}{b_1}{a_1}{b_2}\ C_{b_2}(\l{2})\
C_{b_1}(\l{1})\quad ,\cr}}

\equation{r2}{\eqalign{\sr{\mu}{a}{b}{c}{d}
&= b({\mu})I^{ab}_{cd} + a({\mu}){\pbf}^{ab}_{cd}\quad ,\cr
\srfb{\mu}{a}{b}{c}{d}
&= b({\mu})I^{ab}_{cd} + a({\mu}){\pfb}^{ab}_{cd}\quad ,\cr}}

\equation{Dstate2}{\eqalign{
D(\mu )|\l{1},\ldots , \l{n}|F\rangle &= \prod_{j=1}^n
{1\over \a{\mu}{\l{j}}} \left({a(\mu )\over a(-\mu )}\right)^L
|\l{1},\ldots , \l{n}|F\rangle\cr
&+ \sum_{k=1}^n\left({\tilde\Lambda}_k\right)^{b_1\ldots
b_n}_{a_1\ldots a_n} C_{b_k}(\mu ) \prod_{\scr j=1\atop \scr
j\neq k}^n C_{b_j}(\l{j})\vac F^{a_n\ldots a_1}\quad ,\cr}}

\equation{Astate2}{\eqalign{
(A_{11}(\mu )-A_{22}(\mu ))|\l{1},&\ldots ,
\l{n}|F\rangle =\cr
&=\ (a(\mu ))^L\prod_{j=1}^n {1\over \a{\mu}{\l{j}}}
\prod_{l=1}^n C_{b_l}(\l{l})\vac\ \tau^{(1)}(\mu )^{b_1\ldots
b_n}_{a_1\ldots a_n}\ F^{a_n\ldots a_1} \cr
&\quad +\sum_{k=1}^n \left(\Lambda_k\right)^{b_1\ldots b_n}_{a_1\ldots a_n}
C_{b_k}(\mu )
\prod_{\scr j=1\atop \scr j\neq k}^n
C_{b_j}(\l{j})\vac F^{a_n\ldots a_1}\quad ,\cr}}

\equation{tausuth}{\eqalign{\tau^{(1)}(\mu )^{b_1\ldots
b_n}_{a_1\ldots a_n} &= (-1)^{\ge_c}\
\Le1{n}{\mu}^{cc_{n-1}}_{b_na_n}
\Le1{n-1}{\mu}^{c_{n-1}c_{n-2}}_{b_{n-1}a_{n-1}}
\ldots\Le1{1}{\mu}^{c_1c}_{b_1a_1}\times\cr
&\qquad\times (-1)^{\ge_c\sum_{i=1}^{n-1}(\ge_{b_i}+1) +
\sum_{i=1}^{n-1}\ge_{c_i}(\ge_{b_i}+1)}\quad .\cr}}

\equation{newgrtp}{\left(F\osl G\right)^{ab}_{cd} = F_{ab}G_{cd}\
(-1)^{(\ge_c +1)(\ge_a+\ge_b)} \quad .}

\equation{tau12}{\eqalign{\tau^{(1)}(\mu )^{b_1\ldots
b_n}_{a_1\ldots a_n} &= str( T_n^{(1)}(\mu ))=
str(\Le1{n}{\mu}\osl\Le1{n-1}{\mu}\osl\ldots
\osl\Le1{1}{\mu})\ ,\cr \noalign{\vskip 4pt}
L_k^{(1)}(\la) &= b(\la) \pbf + a(\la) I^{(1)}\ =
\left(\matrix{a(\la)+b(\la)\ e_k^{11}&b(\la)\ e_k^{21}\cr b(\la)\
e_k^{12}&a(\la)-b(\la)e_k^{22}\cr}\right)\quad .\cr}}

\equation{cancelbff}{\left((\Lambda_k)^{b_1\ldots b_n}_{a_1\ldots a_n} -
(\tilde\Lambda_k)^{b_1\ldots b_n}_{a_1\ldots a_n}\right)
F^{a_n\ldots a_1} = 0\quad }

\equation{pbcbff1}{F^{a_n\ldots a_1} = \left(a(-\l{k})\right)^L
\left(\tau^{(1)}(\l{k})F\right)^{a_n\ldots a_1}\ \ ,\ \ k=1,\ldots,
n\quad .}

\equation{otherr}{{{\hat r}(\mu )}^{ab}_{cd} = b(\mu ) \gd_{ab}\ \gd_{cd} +
a(\mu ) \gd_{ad}\ \gd_{bc}\ (-1)^{\ge_a +\ge_c + \ge_a\ge_c}\quad .}

\equation{vacnest2}{{\vac ^{(1)}_k = \pmatrix{0\cr 1\cr}\qquad ,\qquad
\vac ^{(1)} = \osl_{k=1}^n \vac ^{(1)}_k}\quad }

\equation{intnestbff1}{{\hat r}(\l{}-\mu)\
T_n^{(1)}(\l{})\osl T_n^{(1)}(\mu)=T_n^{(1)}(\mu)\osl
T_n^{(1)}(\l{})\ {\hat r}(\l{}-\mu)}

\equation{intnestbff2}{{r}(\l{}-\mu)\
T_n^{(1)}(\l{})\otimes T_n^{(1)}(\mu)=T_n^{(1)}(\mu)\otimes
T_n^{(1)}(\l{})\ {r}(\l{}-\mu)\quad .}

\equation{BE2}{\eqalign{\left(a(-\l{l})\right)^L &= \prod_{\scr
m=1\atop \scr m\neq l}^n {a(\l{m} - \l{l})\over a(\l{l} -
\l{m})}\ \prod_{j=1}^{n_1} a(\l{l} - \1l{j})\quad l=1,\ldots ,n\cr
1&=\prod_{j=1}^n a(\l{j} -\1l{k})\qquad\qquad\qquad\qquad k=1,\ldots ,
n_1\ \ .\cr}}

\equation{shift}{\tl{j}=\l{j} - {i\over 2}\ \ ,\qquad
\2t{j} = \1l{j} -i\quad ,}

\equation{BE22}{\eqalign{\left(\tb{\tl{l}}\right)^L &= \prod_{\scr
m=1\atop \scr m\neq l}^{N_h+N_\downarrow} {\tc{\tl{l} - \tl{m}}}\
\prod_{j=1}^{N_h} \ta{\tl{l} - \2t{j}}\quad l=1,\ldots
,N_h+N_\downarrow\cr
1&=\prod_{j=1}^{N_h+N_\downarrow} \ta{\tl{j}
-\2t{k}}\qquad\qquad\qquad\qquad k=1,\ldots , N_h\ \ .\cr}}

\equation{taubff}{\eqalign{\nu (\mu,\l{1},\ldots , \l{n}) &=
(a(\mu ))^L\prod_{j=1}^{N_h+N_{\downarrow}} {1\over \a{\mu}{\l{j}}} \nu
^{(1)}(\mu )\ - \ \prod_{j=1}^{N_h+N_{\downarrow}} {1\over
\a{\mu}{\l{j}}}\left({a(\mu )\over a(-\mu )}\right)^L\cr
\nu ^{(1)}(\mu ) &=
\prod_{i=1}^{N_h}{1\over a(\mu - \1l{i})}
\left(\prod_{j=1}^{N_h+N_{\downarrow}}\a{\mu}{\l{j}}-
\prod_{j=1}^{N_h+N_{\downarrow}}
{\a{\mu}{\l{j}}\over \a{\l{j}}{\mu}}\right)\quad .\cr}}

\equation{ebff}{E_{susy} = L-\sum_{j=1}^{N_h+N_{\downarrow}} {1\over
{(\tl{j})}^2 +{1\over 4}}=
L-2(N_h+N_{\downarrow})-2\sum_{j=1}^{N_h+N_{\downarrow}}cos(k_j)\quad
,}


\equation{BE3}{\eqalign{\left(a(-\l{l})\right)^L &=
\prod_{j=1}^{N_{\downarrow}} a(\1l{j} - \l{l})\qquad\quad l=1,\ldots
,N_h+N_{\downarrow}\cr 1&=\prod_{j=1}^{N_h+N_{\downarrow}} a(\1l{k}
-\l{j})\qquad k=1,\ldots , N_{\downarrow}\ \ .\cr}}

\equation{shift2}{\tl{j}=\l{j}-{i\over 2}\quad ,}

\equation{BE32}{\eqalign{\left(\tb{\tl{l}}\right)^L &=
\prod_{j=1}^{N_{\downarrow}} \ta{\1l{j} - \tl{l}}\qquad\quad l=1,\ldots
,N_h+N_{\downarrow}\cr
1&=\prod_{j=1}^{N_h+N_{\downarrow}} \ta{\1l{k}
-\tl{j}}\qquad k=1,\ldots , N_{\downarrow}\ \ .\cr}}

\equation{taufbf}{\eqalign{\nu (\mu,\lbrace\l{j}\rbrace, F) &=
(a(\mu ))^L\prod_{j=1}^{N_h+N_{\downarrow}} {1\over \a{\mu}{\l{j}}}
\left(\nu ^{(1)}(\mu )\ - \ \left({1\over a(-\mu )}\right)^L\right)\cr
\nu ^{(1)}(\mu ) &=
\prod_{i=1}^{N_{\downarrow}}{1\over a(\1l{i}-\mu )}
\left(1-\prod_{j=1}^{N_h+N_{\downarrow}} \a{\mu}{\l{j}}\right)\quad .\cr}}

\equation{efbf}{E_{susy} = L-\sum_{j=1}^{N_h+N_{\downarrow}} {1\over
(\tl{j})^2 +{1\over 4}\quad }.}

\equation{strings}{\eqalign{{\tilde\la}_{n,j} &= \l{n}+{i\over 2}
(n+1-2j)\ ,\quad j=1\ldots n\cr
\la^{(1)}_{n-1,k} &= \l{n}+{i\over 2} (n-2k)\ ,\quad k=1\ldots
n-1\quad .\cr}}

\equation{BAEnew}{\eqalign{\left(\tb{\tl{l}}\right)^L &=
\prod_{j=1}^{N_{\downarrow}} \ta{\1l{j} - \tl{l}}\qquad\quad l=1,\ldots
,N_h+N_{\downarrow}\cr
1&=\prod_{j=1}^{N_h+N_{\downarrow}} \ta{\1l{k}
-\tl{j}}\qquad k=1,\ldots , N_{\downarrow}\ \ \cr}}

\equation{pol}{p(w)=\prod_{j=1}^{N_h+N_{\downarrow}} (w-\tl{j}-{i\over
2}) - \prod_{j=1}^{N_h+N_{\downarrow}} (w-\tl{j}+{i\over 2}) = 0\quad
.}

\equation{res}{\sum_{j=1}^{N_\da} -i\log\left({{\tilde\la}_l-w_j+{i\over
2}\over{\tilde\la}_l-w_j-{i\over 2}}\right) =
\sum_{j=1}^{N_\da}{1\over 2\pi i}\oint_{C_j} dz
(-i)\log\left({{\tilde\la}_l-z+{i\over 2}\over{\tilde\la}_l-z-{i\over
2}}\right)\ {d\over dz}\log(p(z))\quad ,}

\equation{res2}{\sum_{j=1}^{N_\da}
-i\log\left({{\tilde\la}_l-w_j+{i\over
2}\over{\tilde\la}_l-w_j+{i\over 2}}\right) =  -\sum_{j=1}^{N_h}
-i\log\left({{\tilde\la}_l-{w^\prime}_j+{i\over
2}\over{\tilde\la}_l-{w^\prime}_j-{i\over
2}}\right)-i\log\left({p(z_n)\over p(z_p)}\right)\quad ,}

\equation{cut}{\eqalign{p(z_n) &= -
\prod_{m=1}^{N_h+N_\downarrow}\left(\tl{l}-\tl{m}+i \right)\cr
p(z_p) &=
\prod_{m=1}^{N_h+N_\downarrow}\left(\tl{l}-\tl{m}-i \right)\quad
.\cr}}

\equation{id}{\prod_{j=1}^{N_\da}{\tl{l}-w_j+{i\over 2}\over
\tl{l}-w_j-{i\over 2}} =
\prod_{k=1}^{N_h}{\tl{l}-{w^\prime}_k-{i\over 2}\over
\tl{l}-{w^\prime}_k+{i\over 2}}\
\prod_{\scr m=1\atop \scr m\neq l}^{N_h+N_\downarrow} {\tc{\tl{l} -
\tl{m}}}\quad .}

\equation{BAEsuth1}{\left(\tb{\tl{l}}\right)^L = \prod_{\scr
m=1\atop \scr m\neq l}^{N_h+N_\downarrow} {\tc{\tl{l} - \tl{m}}}\
\prod_{k=1}^{N_h}{\tl{l}-{w^\prime}_k-{i\over 2}\over
\tl{l}-{w^\prime}_k+{i\over 2}}\
\quad l=1,\ldots ,N_h+N_\downarrow\quad .}

\pagenumstyle{blank}
\footnoteskip=2pt
\line{\it March 1992\hfil ITP-SB-92-12}

\vskip4em
\baselineskip=32pt
\begin{center}{\bigsize{\sc A new solution of the supersymmetric t-J
model\\
by means of the Quantum Inverse Scattering Method}
{\baselineskip=20pt\footnote{This work was supported in part by
the National Science Foundation under research grants PHY91-07261 and
NSF91-08054.}}}
\end{center}
\vfil
\baselineskip=16pt

\begin{center}
{\bigsize
Fabian H.L.E\sharps ler\footnote[$\ \flat$]{\sc e-mail:
fabman@max.physics.sunysb.edu}\vskip .5cm
and \vskip .5cm
Vladimir E. Korepin\footnote[$\ \sharp$]{\sc e-mail:
korepin@dirac.physics.sunysb.edu}\vskip .5cm}

\vskip 1cm

\it Institute for Theoretical Physics\vskip 4pt
State University of New York at Stony Brook\vskip 4pt
Stony Brook, NY~~11794-3840

\end{center}

\vfil

\centertext{\bfs \bigsize ABSTRACT}
\vskip\belowsectionskip

\begin{narrow}[4em]
We construct the enveloping fundamental spin model of the t-J
hamiltonian using the Quantum Inverse Scattering Method (QISM),
and present all three possible Algebraic Bethe Ans\"atze.
Two of the solutions have been previously obtained in the framework of
Coordinate Space Bethe Ansatz by Sutherland and by Schlottmann and
Lai, whereas the third solution is new.
The formulation of the model in terms of the QISM enables us to
derive explicit expressions for higher conservation laws.

\end{narrow}
\vskip 1cm
{\sc PACS : 75.10.Jm\ \ 71.20.Ad}
\vfil

\break


\footnotenumstyle{arabic}
\sectionstyle{left}
\pagenumstyle{arabic}
\pagenum=0

{\bfs\section{Introduction}}
\sectionnum=1
Strongly correlated electronic systems are currently intensely studied
in relation with high $T_c$ superconductivity. Recently there has been
a renewed interest in the one-dimensional t-J model as an integrable
low dimensional version of a strongly correlated electronic system.
The t-J model was proposed by F.C. Zhang and T.M. Rice\upref zhri/.
P.W. Anderson claimed that two-dimensional systems may share features
of one-dimensional systems\upref pw2/, which could imply a certain
relevance of some results obtained for the one-dimensional t-J model
to high $T_c$ superconductivity.

The model describes electrons on a one-dimensional lattice with a
Hamiltonian that includes nearest neighbour hopping (t) and nearest
neighbour spin exchange and charge interactions (J). The Hilbert space
of the model is constrained to exclude double occupancy of any single
site, which corresponds to an infinite on-site repulsion.

Electrons on a lattice are described by operators $c_{j,\sigma}\ $,
$j=1,\ldots,L$, $\sigma=\pm 1$, where $L$ is the total number
of lattice sites. These are canonical Fermi operators with
anti-commutation relations given by $\{ c^\dagger_{i,\sigma} ,
c_{j,\tau} \} = \delta_{i,j} \delta_{\sigma,\tau}$. The state
$\vac $ (the Fock vacuum) satisfies $c_{i,\sigma} \vac = 0$.
Due to the constraint excluding double occupancy there are three
possible electronic states at a given lattice site $i$
$$\putequation{config}$$
By $n_{i,\sigma}= c^\dagger_{i,\sigma} c_{i,\sigma}$ we denote the number
operator for electrons with spin $\sigma$ on site $i$ and we write
$n_i=n_{i,1} + n_{i,-1}$. The spin-operators $S=\sum_{j=1}^L
S_j$, $S^\dagger =\sum_{j=1}^L S^\dagger_j$, $S^z=\sum_{j=1}^L S^z_j$,
$$\putequation{spin}$$
form an $su(2)$ algebra and they commute with the hamiltonian
that we consider below. [We shall always give local expressions
${\cal O}_j$ for symmetry generators, implying that the global ones
are obtained as ${\cal O} = \sum_{j=1}^L {\cal O}_j$.]
The Hamiltonian on a lattice of $L$ sites is given by the following
expression :
$$\putequation{h}$$
The projectors $(1-n_{j,-\sigma})$ in the kinetic energy terms ensure
that $H$ acts within the constrained Hilbert space, {\sl i.e.} does not
create states with double occupancy.
A equivalent expression for $H$ is
$$\putequation{hugly}$$
where ${\cal P}$ is the projector on the subspace of non-doubly occupied
states.
It was shown in [{\putref{Schlotti}] that the model described by
(\putlab{h}) is integrable and can be solved by coordinate space Bethe
Ansatz. In addition it is possible to map it onto the integrable
quantum lattice gas of Lai and Sutherland{\upref Lai,Bill/}.

The thermodynamics of the model were treated in [\putref{Schlotti,
Sarkar}] and the ground state properties and excitation spectrum were
investigated in [\putref{BarBlO,BarBl,Bares}].

As the number operator for electrons $\hat N = \sum_{j=1}^L
\sum_{\sigma = \pm 1} n_{j,\sigma}$ commutes with $H$, we can add a
term $2\ \hat N-L$ to the hamiltonian without changing the set of
eigenvectors.
Physically this just amounts to a shift of the chemical potential.
For the special value\footnote{The case $-J=2t$ can be
obtained from $J=2t$ {\sl via} the transformation \hfill\break
$c_j\rightarrow (-1)^j c_j\ ,\quad c^\dagger_j\rightarrow (-1)^j
c^\dagger_j$} $J=2t=2$ the shifted hamiltonian now exhibits a number
of interesting properties:\hfill\break
Firstly it is supersymmetric, {\sl i.e.} it commutes with all nine
generators of the superalgebra $u(1|2)$\upref Sarkar,BarBlO/.
Secondly it can be written as a graded permutation operator\upref
Sarkar,BarBlO/:
$$\putequation{hpi}$$
The operator $\Pg ^{j,j+1}$ permutes
the three possible configurations (\putlab{config})
between the sites $j$ and $j+1$, picking up a minus sign if both of
the permuted configurations are fermionic, {\sl i.e.}
$$\putequation{perm}$$
It is clear that this form of interaction conserves the individual
numbers $N_\uparrow$ and $N_\downarrow$ of electrons with spin up
and spin down, and due to the constraint of no double
occupancy the number $N_h$ of empty sites (or ``holes'') is also
conserved.\vskip .5cm

The outline of the paper is as follows :\hfill\break
In section $2$ we give a discussion of the supersymmetry of the model
and express the hamiltonian (\putlab{hpi}) in a way most suitable for
the analysis of section $5$.
In section $3$ we perform a detailed construction of the
Algebraic Bethe Ansatz of the model.
We derive three different forms for the Bethe Ansatz Equations (BAE)
and the eigenvalues of the transfer matrix.
Two of these solutions have previously been obtained by
various authors\upref Schlotti,Lai,Bill,Kul/, whereas our third
solution, presented in section $4$, is new.
Our expression for the BAE seems to be
particularly simple and we hope that it will be useful in clarifying
the physical features of the model. The direct physical
consequences of our new solution (like the structure of the ground
states and classification of excitations and correlations) are
currently under investigation\upref{ek}/ and will presented in a
future publication.
The graded Quantum Inverse Scattering Method (QISM), discussed in
section $3$, enables us to obtain expressions for (an infinite
number of) higher conservation laws at the quantum level. These
conserved charges are of interest, because physical interactions,
although of short range, are not generally well approximated by
interactions involving only nearest neighbours.
The charges under consideration involve interactions of longer range
(next nearest neighbours, next next nearest {\sl etc.})
and can be added to the hamiltonian without spoiling the
integrability of the model. Thus it is possible to construct
integrable models with longer range interactions by using higher
conservation laws\upref Frahm/.
The first nontrivial higher integral of motion is for example given by
the expression
$$\putequation{hintro}$$
where $H^{k,k+1}$ is the density of the hamiltonian defined in
(\putlab{hpi}).
Section $5$ is devoted to the derivation of explicit formulas for
higher conservation laws.
{\bfs \section{Supersymmetry of the t-J model}}
For $J=2t$ the t-J model exhibits a (global) $u(1|2)$ invariance
on the constrained Hilbert space. In the literature this symmetry
algebra has frequently been denoted by $spl(2,1)$\upref BarBlO/. The relation
between $spl(2,1)$ and $su(1|2)$ (neglecting the trivial $u(1)$ factor
for the time being) is as follows :\hfill\break
The algebra $su(1|2)$ is a real form of the complex Lie-superalgebra
$sl(1,2;{\Complex})$, whereas $spl(2,1)$ is an equivalent notation for
$sl(1,2;{\Complex})$. For our purposes it is more convenient to work
with a real base field, so that we will work in a representation of
$u(1|2)$.
The generators of the $u(1|2)$ algebra are given by \upref sup/ (we
write ${\cal O} = \sum_{j=1}^L {\cal O}_j$)

$$\putequation{u12}$$

The operators $S,\ S^\dagger,\ S^z,\ Q_1,\ Q^\dagger_1,\ Q_{-1},\
Q^\dagger_{-1},\ T$ generate the $su(1|2)$ subalgebra of $u(1|2)$.

In the fundamental representation there exists an invariant,
nondegenerate bilinear form $K_{\alpha\beta}$ on $u(1|2)$, which is
given as the supertrace over two generators \upref Cornwell/
$$\putequation{kab}$$
and which is explicitly computed in Appendix A.
Note that the nondegeneracy of $K_{\ga\gb}$ is a feature of the
fundamental representation and does not generally hold for other
representations, because $u(1|2)$ is not semisimple.
For later use we define the structure constants of $u(1|2)$ as
$$\putequation{f}$$
where $\ge_\ga$ are the Grassmann parities of the generators
$J_{k,\ga}$ ({\sl i.e. }$\ge =1$ for the fermionic generators $J_4\ldots
J_7$ and $\ge=0$ for the rest).

The Hamiltonian of the t-J model on the constrained Hilbert space can
now be expressed in terms of the densities $J_{k,\ga}$
as
$$\putequation{H}$$
where we have used (\putlab{u12}) and the explicit expression for
$K_{\ga\gb}$ given in Appendix A.
In this form the global $u(1|2)$ invariance of the hamiltonian is
easily confirmed
$$\putequation{inv}$$
The ``manifestly supersymmetric'' expression for the hamiltonian
(\putlab{H}) in terms of the Killing form will enable us to express
the higher conservation laws we derive in section $4$, in an $u(1|2)$
invariant way.
{\bfs\section{Graded Quantum Inverse Scattering Method}}
In this section we construct the enveloping spin model of the
hamiltonian of the one-dimensional supersymmetric t-J model, using
the Quantum Inverse Scattering Method (QISM).
Due to the fact that we are dealing with a supersymmetric theory it is
necessary to modify the QISM along the lines discussed
in [\putref{Kul, SklKul}].
Below we give a summary of the ``graded'' version of the QISM.
We start with an $R$-matrix, obeying a graded Yang Baxter
equation, and construct from it a ``fundamental'' spin model ({\sl
i.e.} the $L$-operator is constructed directly from the $R$-matrix).
We then show that the trace identities of the corresponding
transfer matrix give rise to the hamiltonian of the t-J model.
Finally we construct a set of simultaneous eigenstates of the
transfer matrix and the hamiltonian, using a nested Algebraic Bethe
Ansatz\footnote{For a general introduction to the Algebraic
Bethe Ansatz we refer to the publications
(\putref{Vlad,FT,FT2}).}(NABA)\upref Kul,KulRe,KulRe2,Taktajan/.
Due to the grading there exist three choices for the $R$-matrix, all
of them describing the same physical system, but leading to different
(yet equivalent as shown in Appendix C) forms of the NABA.
Two of these possibilities of performing a Bethe
Ansatz analysis are equivalent to the coordinate space Bethe Ansatz
solutions of Lai and Sutherland. We reproduce their respective
periodic boundary conditions and expressions for energy eigenvalues of
the hamiltonian.

\subsection{3.1.\ \sl Yang Baxter Equation}

Consider the graded linear space $V^{(m|n)}= V^m \oplus V^n $, where
$m$ and $n$ denote the dimensions of the ``even'' ($V^m$) and ``odd''
($V^n$) parts, and $\oplus$ denotes the direct sum.
Let $\lbrace e_1,\ldots ,e_{m+n}\rbrace$ be a basis of $V^{(m+n)}$,
such that $\lbrace e_1,\ldots ,e_m\rbrace$ is a basis of $V^m$ and
$\lbrace e_{m+1},\ldots ,e_{m+n}\rbrace$ is a basis of $V^n$.
The Grassmann parities of the basis vectors are given by $\ge_1=\ldots
=\ge_m =0$ and $\ge_{m+1}=\ldots =\ge_{m+n} =1$.
Linear operators on $\vmn$ can be represented in block form ($M\in
End(\vmn)$)
$$\putequation{linops}$$
The supertrace is defined as
$$\putequation{str}$$
where the traces on the r.h.s. are the usual operator traces in $V^m$
and $V^n$.
We now define the graded tensor product space $V^{(m|n)}\otimes
V^{(m|n)}$ in terms of its basis vectors $\lbrace e_a\otimes
e_b|a,b=1,\ldots, m+n\rbrace$ as follows
$$\putequation{tensorprod}$$
Compared to the ``ordinary'' tensor product the additional factor
$(-1)^{\ge_{v_a}\ge_b}$ occurs, which originates in passing $v_a$ past
$e_b$.
The action of the right linear operator $F\otimes G$ on the vector
$v\otimes w$ in $\vmn\otimes\vmn$ is given by
$$\putequation{linop}$$
Therefore its matrix elements are of the form
$$\putequation{linopmat}$$
The identity operator in $\vmn\otimes\vmn$ is given by
$I^{a_1 b_1}_{a_2 b_2} = \delta_{a_1b_1}\delta_{a_2b_2} $
and the matrix $\Pg$ that permutes the individual linear spaces in the
tensor product space, is of the form
$$\putequation{pigr}$$
The physical relevance of the above construction is as follows :
If we consider a lattice gas of $m$ species of bosons and $n$ species
of fermions, then $V^{(m|n)}$ denotes the space of configurations at
every site of the lattice. For the example of the t-J model we have
$m=1,\ n=2$ and the three allowed configurations are given by
(\putlab{config}). The tensor product space $\v12\otimes\v12$
describes two neighbouring sites, and owing to the fermionic nature of
some of the configurations, the tensor product has to carry a grading.
$\Pg$ permutes configurations on neighbouring
sites, and we pick up a minus sign if we permute two fermions.

A matrix $R(\l{})$ (depending on a spectral parameter $\l{}$) is said
to fulfill a graded Yang-Baxter-equation, if the
following identity on $\vmn\otimes \vmn\otimes \vmn$ holds
$$\putequation{YBE1}$$
In components this identity reads
$$\putequation{YBE}$$
Note that despite the fact that the tensor product in (\putlab{YBE1})
carries a grading, there are no extra signs in (\putlab{YBE}) compared
to the nongraded case.
It is easily checked that the R-matrix
$$\putequation{Rgen}$$
fulfills equation (\putlab{YBE}).

\subsection{3.2.\ \sl Construction of the Transfer matrix}
\subsectionnum=2

By multiplying (\putlab{YBE}) by
$\Pg^{f_3 e_3}_{f_1 a_1}\ \Pg^{e_2 a_2}_{e_3 a_3}$
from the left one can derive the equation
$$\putequation{int}$$
In matrix notation (\putlab{int}) reads
$$\putequation{YBE2}$$
where the indices $1,2,3$ indicate in which of the spaces $\vmn$ in
the tensor product space $\vmn\otimes \vmn\otimes \vmn$ the matrices
act nontrivially.
The tensor product in (\putlab{YBE2}) is between the spaces $1$ and
$2$.  We now call the third space ``quantum space''and the first two
spaces ``matrix spaces''.
The physical interpretation of the quantum space is as the Hilbert
space over a single site of a one-dimensional lattice. We now consider
the situation, where intertwining relations of the type
(\putlab{YBE2}) hold for all sites of a lattice of length $L$. The
quantum space index ``$3$'' now gets replaced by an index labelling
the number of the site.
We define the $L$-operator (on site $n$) as a linear operator on
${\cal H}_n\otimes V^{(m|n)}_{matrix}$ (where ${\cal H}_n\simeq \vmn$
is the Hilbert space over the $n^{th}$ site, and $V^{(m|n)}_{matrix}$ is
a matrix space)
$$\putequation{LR}$$
$L_n$ is a quantum operator valued $(m+n)\times (m+n)$ matrix, with
quantum operators acting nontrivially in the $n^{th}$ quantum space
(of the direct product Hilbert space over the complete lattice
$\otimes_{j=1}^L {\cal H}_j$).
The greek indices are the ``quantum indices'' and the roman indices
are the ``matrix indices''.
Equation (\putlab{YBE2}) for the $n^{th}$ quantum space can now be
rewritten as the operator equation
$$\putequation{int1}$$
In matrix notation (\putlab{int1}) takes the form
$$\putequation{int2}$$
Here the graded tensor product is between the two matrix spaces and
$R$ only acts in the matrix spaces.
The intertwining relation (\putlab{int2}) enables us to construct an
integrable spin model as follows :\hfill\break
We first define the monodromy matrix $T_L(\l{})$ as the matrix product
over the $L$-operators on all sites of the lattice, {\sl i.e.}
$$\putequation{mon}$$
$T_L(\l{})$ is a quantum operator valued $(m+n)\times (m+n)$ matrix
that acts nontrivially in the graded tensor product of all quantum
spaces of the lattice and by construction fullfills the same
intertwining relation as the $L$-operators (as can be proven by
induction over the length of the lattice)
$$\putequation{intT}$$
The transfer matrix $\tau (\l{})$ of the integrable spin model is now
given as the matrix supertrace of the mondromy matrix
$$\putequation{tau}$$
As a consequence of (\putlab{intT}) transfer matrices with different
spectral parameters commute. This condition implies that
the transfer matrix is the generating functional of
the hamiltonian and of an infinite number of ``higher'' conservation
laws of the model.

\subsection{3.3.\ \sl Trace Identities}
\subsectionnum=3

Taking logarithmic derivatives of the transfer matrix at a special
value of the spectral parameter, one can generate higher conservation
laws\upref{Vlad}/.
For our specific case at hand, {\sl i.e.} the $R$-matrix
(\putlab{Rgen}), the corresponding hamiltonian is obtained by taking
the first logarithmic derivative at zero spectral parameter
$$\putequation{trid1}$$
The proof of this identity can be carried out in the same way as for
the ungraded case, the main difference being the grading
of the tensor product of the quantum spaces (see (\putlab{mon})).
By shifting the energy eigenvalues by a constant we
obtain the expression (\putlab{hpi}) for the hamiltonian of the t-J
model
$$\putequation{trid2}$$
if we choose our underlying graded vector space to have signature
$(1,2)$, {\sl i.e.} to have a basis with two fermionic and one bosonic
state.
This shows that the transfer matrix constructed from the $L$-operator
(\putlab{LR}) and $R$-matrix (\putlab{Rgen}) is indeed the correct
transfer matrix for the one-dimensional supersymmetric t-J model.
Higher conservation laws are obtained as the coefficients of the power
series
$$\putequation{hcl}$$
There exists however a simpler method for the construction of higher
integrals of motion than taking logarithmic derivatives, which we will
discuss in section $5$.

\subsection{3.4.\ \sl Algebraic Bethe Ansatz with a bosonic background
(FFB grading)\hfill\break \null\qquad Lai solution}

Due to the constraint of no double occupancy there are three different
configurations per site for the t-J model.
Thus the Hilbert space at the $k^{th}$ site of the lattice is
isomorphic to ${\Complex}^3$ and is spanned by
the three vectors $e_1 = \pmatrix{1&0&0\cr}^T$, $e_2 =
\pmatrix{0&1&0\cr}^T$ and $e_3 = \pmatrix{0&0&1\cr}^T$.
In this section we consider a grading such that
$e_1$ and $e_2$ are fermionic (representing spin down/spin up
electrons respectively) and $e_3$ is bosonic (empty site).
In terms of the Grassmann parities this means that $\ge_1=\ge_2=1$ and
$\ge_3 = 0$.
We pick the reference state in the $k^{th}$ quantum space $\vac _k$,
and the vacuum $\vac$ of the complete lattice of $L$ sites, to
be purely bosonic, {\sl i.e.}
$$\putequation{vacuum}$$
This choice of grading implies that $R(\mu ) =b(\mu ) I + a(\mu ) \Pg$
is given by the following expression
{\baselineskip=12pt $$\putequation{R}$$}
The $L$-operator is defined by (\putlab{LR}) and is of the form
$$\putequation{Lop}$$
where $e_n^{ab}$ are quantum operators in the $n^{th}$ quantum space
with matrix representation $(e_n^{ab})_{\ga\gb} = \gd _{a\ga}\gd
_{b\gb}$.
The monodromy matrix (\putlab{mon}) is a quantum operator valued
$3\times 3$ matrix, which we represent as
$$\putequation{T}$$
The transfer matrix is then given as
$$\putequation{tau2}$$
The action of $L_k(\l{})$ on the reference state on the $k^{th}$ site
is
$$\putequation{Lvac}$$
Using (\putlab{mon}) and (\putlab{Lvac}) we determine the action of
the monodromy matrix on the reference state to be
$$\putequation{Tvac}$$
We will now construct a set of eigenstates of the transfer matrix
using the technique of the NABA.
Inspection of (\putlab{Tvac})
reveals that $C_1(\la )$ and $C_2(\la )$ are creation operators (of
odd Grassmann parity) with respect to our choice of reference state.
This observation leads us to the following Ansatz for the eigenstates
of $\tau(\mu )$
$$\putequation{state1}$$
where the indices $a_j$ run over the values $1,2$, and
$F^{a_n\ldots a_1}$ is a function of the spectral parameters $\l{j}$.
The action of the transfer matrix on states of the form
(\putlab{state1}) is determined by (\putlab{Tvac}) and the
intertwining relations (\putlab{intT}).
The components of the intertwining relations (\putlab{intT}) needed
for the construction of the NABA are
$$\putequation{AC}$$
where
$$\putequation{r}$$
Here ${\P^{(1)}}^{ab}_{cd} = -\gd_{ad}\ \gd_{bc}$ is the
$4\times 4$ permutation matrix corresponding to the grading
$\ge_1=\ge_2=1$.
$r(\mu)$ can be seen to fulfill a (graded) Yang-Baxter equation on its
own
$$\putequation{ybe}$$
and can be identified with the $R$-matrix of a fundamental spin model
describing two species of fermions.
Using (\putlab{AC}) we find that
the diagonal elements of the monodromy matrix act on the states
(\putlab{state1}) as follows
$$\putequation{Dstate}$$
$$\putequation{Astate}$$
where
$$\putequation{tau1}$$
$L^{(1)}$ and $r(\mu)$ can be
interpreted as $L$-operator and $R$-matrix of a fundamental spin model
($r$ fullfills the Yang-Baxter equation (\putlab{ybe})), describing two
species of fermions. Hence $T_n^{(1)}(\mu )$ and $\tau^{(1)}(\mu )$
are the monodromy matrix and transfer matrix of the corresponding
{\sl inhomogeneous} model.
Inspection of (\putlab{Dstate}) and (\putlab{Astate}) together with
(\putlab{tau2}) shows that the eigenvalue condition
$$\putequation{ew}$$
leads to the requirements that $F$ ought to be an eigenvector of the
``nested'' transfer matrix $\tau^{(1)}(\mu )$, and that the ``unwanted
terms'' cancel, {\sl i.e.}
$$\putequation{unw}$$
The relative sign in (\putlab{unw}) is due to the supertrace in
(\putlab{tau2}) and (\putlab{ew}).
The quantities $\Lambda_k$ and $\tilde\Lambda_k$ are computed in
Appendix A. Using their explicit expressions in (\putlab{unw}) we
obtain the following conditions on the spectral parameters
$\l{j}$ and coefficients $F$, which are necessary for (\putlab{ew}) to
hold
$$\putequation{PBC}$$
This completes the first step of the NABA. In the next step we will
now solve the nesting.
The condition that $F$ ought to be an eigenvector of $\tau ^{(1)}(\mu )$
requires the diagonalisation of $\tau ^{(1)}(\mu )$, which can be carried
out by a second, ``nested'' Bethe Ansatz.
{}From (\putlab{ybe}) and (\putlab{tau1}) the following intertwining
relation is easily derived
$$\putequation{intnest}$$
If we write
$$\putequation{T1}$$
then (\putlab{intnest}) and (\putlab{r}) imply that
$$\putequation{intnest2}$$
As the reference state for the nesting we pick
$$\putequation{vacnest}$$
The action of the nested monodromy matrix $T^{(1)}_n(\mu )$on the
reference state $\vacn $ is determined by (\putlab{tau1}) and we find
$$\putequation{tvac}$$
We now make the following Ansatz for the eigenstates of
$\tau^{(1)}(\mu )$
$$\putequation{state2}$$
These states can be related to the coefficients $F^{a_n\ldots
a_1}$ in the following way :\hfill\break
The state ${|\1l{1},\ldots , \1l{n_1}\rangle}$ ``lives'' on a lattice
of $n$ sites and is thus an element of a direct product over $n$
Hilbert spaces. In components it reads ${|\1l{1}\ldots
\1l{n_1}\rangle}_{a_n\ldots a_1}$, which can be directly identified
with $F^{a_n,\ldots ,a_1}$.

The action of $\tau^{(1)}(\mu )$ on the states (\putlab{state2}) can
be evaluated with the help of the intertwiners (\putlab{intnest2})
$$\putequation{ADstate1}$$
$$\putequation{ADstate2}$$

{}From (\putlab{ADstate1}) and (\putlab{ADstate2})
one can read off the eigenvalues of
$\tau^{(1)}(\mu )$
$$\putequation{eig}$$

Inserting this expression for the special value $\mu = \l{k}$ into
(\putlab{PBC}), we obtain the first set of Bethe equations
$$\putequation{PBC2}$$

The unwanted terms $\Lambda^{(1)}_k$ and $\tilde\Lambda^{(1)}_k$ are
computed in Appendix A and their cancellation (which ensures that the
states (\putlab{state2}) are eigenstates of the transfer matrix
$\tau^{(1)}(\mu )$ ) leads to the following set of Bethe equations for
the nesting
$$\putequation{pbc}$$

Due to our choice of grading $n$ and $n_1$ can be identified as the
total number of electrons and the number of spin down electrons
respectively, {\sl i.e.} $n=N_e=N_{\uparrow} +N_{\downarrow}$, and
$n_1 = N_{\downarrow}$.
If we define shifted spectral parameters according to
${\tilde\l{k}} = \l{k}+{i\over 2}$, we obtain
the Bethe equations in their ``generic'' form
$$\putequation{BE}$$
The eigenvalues of the transfer matrix (\putlab{ew}) are given by
$$\putequation{evtau}$$
Using the trace identities (\putlab{trid2}) it is possible to obtain
the energy eigenvalues from the eigenvalues of the transfer matrix and
we find
$$\putequation{energy}$$
where we have reparametrised ${\tilde\la}_j={1\over 2}cot({k_j\over 2})$.
The periodic boundary conditions (\putlab{BE}) and the energy
(\putlab{energy}) are in perfect agreement with the expressions
derived by Lai\upref Lai/ and by Schlottmann\upref Schlotti/.

\subsection{3.5.\ \sl Algebraic Bethe Ansatz with a fermionic background
I (BFF grading)\hfill\break \null\qquad Sutherland solution}

In this section we consider a grading such that
$e_2$ and $e_3$ are fermionic (representing spin down/spin up
electrons respectively) and $e_1$ is bosonic (empty site).
In terms of the Grassmann parities this means that $\ge_2=\ge_3=1$ and
$\ge_1 = 0$.
We pick the reference state in the $k^{th}$ quantum space $\vac _k$
and the vacuum $\vac$ of the complete lattice of $L$ sites to
be fermionic with all spins up, {\sl i.e.}
$$\putequation{vacuum2}$$
This choice of grading implies that $R$ is of the form
{\baselineskip=12pt $$\putequation{R2}$$}
The L-operator is
$$\putequation{Lop2}$$
The action of $L_k(\l{})$ on the reference state on site $k$ is
$$\putequation{Lvac2}$$
We partition the monodromy matrix as before in (\putlab{T}),
which implies that the transfer matrix is now given by
$$\putequation{tau22}$$
The action of the monodromy matrix on the reference state follows from
(\putlab{Lvac2})
$$\putequation{Tvac2}$$
and inspection of (\putlab{Tvac2}) reveals that $C_{1}(\la )$ and
$C_{2}(\la )$ are creation
operators of odd and even Grassmann parity respectively.
We make the following Ansatz for the eigenstates of $\tau(\mu )$
$$\putequation{state12}$$
The intertwining relations are found to be
$$\putequation{AC2}$$
where
$$\putequation{r2}$$
and $\pbf$ and $\pfb$ are the permutation matrices for the gradings
$\ge_1 = 0\ , \ \ge_2 =1$ and $\ge_1 =1\ ,\ \ge_2 =0$ respectively.
Using (\putlab{AC2}) we find that
the diagonal elements of the monodromy matrix act on the states
(\putlab{state12}) as follows
$$\putequation{Dstate2}$$
$$\putequation{Astate2}$$
where
$$\putequation{tausuth}$$
Here all indices $c_i$ and $c$ are summed over.
The expression for $\tau^{(1)}(\mu )$ is significantly different from
the one in the $FFB$ case treated in section $3.4.$, but
$\tau^{(1)}(\mu)$ can nonetheless be interpreted as the transfer
matrix of an inhomogeneous spin model on a lattice of $n$ sites. Our
reference state $\vac$ is now of fermionic nature and we have to define a
new graded tensor product reflecting this fact
$$\putequation{newgrtp}$$
Effectively the new graded tensor product switches even and odd
Grassmann parities, {\sl i.e.} $\ge_a\longrightarrow \ge_a +1$.
In terms of this tensor product the transfer matrix $\tau^{(1)}(\mu )$
given by (\putlab{tausuth}) can be obtained as
$$\putequation{tau12}$$
In the second line of (\putlab{tau12}) we have explicitly written the
tensor product $\osl$ between the quantum spaces over the sites of the
inhomogeneous model (the $L$-operators are of course again multiplied
as matrices). As before $F^{a_n\ldots a_1}$ must be an eigenvector of
$\tau^{(1)}(\mu )$, if $|\l{1}\ldots \l{n}|F\rangle$ is to be an
eigenstate of $\tau (\mu)$.
The unwanted terms can be computed in a similar way to the ones
described for the $FFB$ grading in Appendix B.
The condition of the cancellation of the unwanted terms
$$\putequation{cancelbff}$$
leads to the conditions
$$\putequation{pbcbff1}$$
To solve the nesting we first have to note that because of our change
of tensor product, the L-operators $L^{(1)}(\la)$ are not intertwined
by the R-matrix $r(\mu)$ defined in (\putlab{r2}), but by the R-matrix
$$\putequation{otherr}$$
The intertwining relation
$$\putequation{intnestbff1}$$
together with the choice of vacuum
$$\putequation{vacnest2}$$
can be analysed along the same lines as for the nesting in section
$3.4$ . It can be shown that they represent a model of the permutation
type with $BF$ grading (describing one species of bosons and one
species of fermions).
The resulting Bethe equations are
$$\putequation{BE2}$$
The choice of grading implies that $n$ and $n_1$ are the number of
holes plus the number of spins down and the number of holes
respectively.
If we shift the spectral parameters according to
$$\putequation{shift}$$
we obtain Sutherland's form of the periodic boundary
conditions\upref{Bill}/ \footnote{For an odd number of lattice sites
Sutherland's equation $(62\ga)$ should be corrected by a factor of
$-1$.}
$$\putequation{BE22}$$
The eigenvalues of the transfer matrix are
$$\putequation{taubff}$$
This results in energy eigenvalues
$$\putequation{ebff}$$
where we have reparametrised $\tl{j}={1\over 2} tan({k_j\over 2})$.

{\bfs\section{A new solution of the t-J model (FBF grading)}}
The third and last possibility of choosing the grading is $\ge_1 =
\ge_3 =1,\ \ge_2 = 0$, with $e_1$ representing spin down and $e_3$
spin up.
This case can be analysed in precisely the same way as
the BFF case so that we simply give the final results for the Bethe
equations and eigenvalues of the transfer matrix. The BAE are
$$\putequation{BE3}$$
We again shift the spectral parameters
$$\putequation{shift2}$$
to obtain the final expression for the new set of periodic boundary
conditions for the t-J model
$$\putequation{BE32}$$
The equivalence of the BAE (\putlab{BE32}) to the BAE (\putlab{BE22})
is demonstrated in Appendix C.
The eigenvalues of the transfer matrix are
$$\putequation{taufbf}$$
The energy eigenvalues can be obtained from (\putlab{taufbf}) in the
usual way by taking logarithmic derivatives
$$\putequation{efbf}$$
The string solutions of the BAE (\putlab{BE32}) are of a very
particular structure :\hfill\break
$n$ spectral parameters ${\tilde\la}_{n,j}$ ($j=1\ldots n$) combine
with $n-1$ spectral parameters $\la^{(1)}_{n-1,k}$ ($k=1\ldots n-1$)
of the nesting into one complex string solution
$$\putequation{strings}$$
Note that due to the symmetry of the hamiltonian under interchange of
spin-up and spin-down there are only three different NABA solutions,
as the other three solutions can be obtained {\sl via} the
substitution $N_\downarrow\leftrightarrow N_\uparrow$.
{\bfs\section{Higher Conservation Laws}}

In this section we derive explicit expressions for the conservation
laws $H_{(3)}$ and $H_{(4)}$ (which involve interactions between $3$
and $4$ neighbouring sites respectively), using a generalisation of
Tetel'man's method\upref Tetelman, Sklyanin/ to the
supersymmetric case.

Let us define the ``boost''-operator
$$\putequation{boost}$$
where $H_{(2)}^{n,n+1}$ is the density of the hamiltonian given by the
right hand side of (\putlab {trid1}).
This operator obviously violates periodicity on the finite chain, but
one can use it nonetheless in commutators which ``differentiate'' the
linear $n$-dependence and yield formally periodic expressions.

The integrals of motion can now be sucessively obtained by commutation
with the boost-operator
$$\putequation{hcl2}$$
where
$$\putequation{boost2}$$

This can be seen as follows:\hfill\break
If we introduce the matrix
$$\putequation{Rtilde}$$
we can rewrite the intertwining relation (\putlab{int2}) as
$$\putequation{rllkul}$$
Here the $L$-operators are multiplied as quantum operators on both
sides of (\putlab{rllkul}).
If we interchange the roles of matrix and quatum spaces in
(\putlab{rllkul}), we obtain the ``$90$ degree rotated'' intertwining
relation
$$\putequation{90rot}$$
where the index ``$1$'' indicates the matrix space (which is the same
for both $L$'s, they are now multiplied as matrices) and $n$ and $n+1$ label
the quantum spaces. The tensor product is now between the quantum
spaces and $R_{n,n+1}$ is a quantum operator acting in both quantum
spaces. From now on we will drop the matrix space index
on the $L$-operators.
In components (\putlab{90rot}) reads
$$\putequation{90rot1}$$
Now we note that (\putlab{Rtilde}) implies that
$$\putequation{diff}$$
Multiplying (\putlab{90rot1}) by $\la -\mu +i$, differentiating with
respect to $\la$, setting $\la = \mu$, and finally acting on both
sides of the equation with
$\left(\Pg^{n,n+1}\right)^{\gb_1\gs_1}_{\gb_2\gs_2}\
(-1)^{\ge_{\ga_1}\ge_{\ga_2}+\ge_{\gs_1}\ge_{\gs_2}}$
from the right, we find
$$\putequation{suth}$$
The tensor product is once again between
quantum space and the dot denotes differentiation with respect to $\mu$.
{}From (\putlab{trid1}) it now follows that
$$\putequation{suth2}$$
Using (\putlab{suth}) it is easy to show that (up to the usual
``problems'' with periodicity)
$$\putequation{hcl-1}$$
and thus
$$\putequation{hcl0}$$
Expanding both sides of (\putlab{hcl0}) in powers of $\mu$ we obtain
(\putlab{hcl2}).
We will now use (\putlab{hcl2}) to obtain explicit
expressions for higher conservation laws.
According to (\putlab{H}) we can write the t-J Hamiltonian in terms of
$u(1|2)$ generators as
$$\putequation{Hagain}$$
$H_{(3)}$ can be obtained by commutation with the boost operator
$\tilde B$
$$\putequation{H3}$$
Using the expressions of the $u(1|2)$ generators
(\putlab{u12}) and the form of $K^{\ga\gb} =
(K_{\ga\gb})^{-1}$ given in Appendix A it is possible to rewrite
$H_{(3)}$ in terms of fermionic creation and annihilation operators
\vfill
$$\putequation{H3again}$$
The $u(1|2)$ generators $Q_{j,\gs},\ Q^\dagger_{j,\gs}$ are given by
(\putlab{u12}).
The next highest conservation law can be computed along similar lines
and we find
$$\putequation{H4}$$
where ${P}^{k-1,k+1}$ is a graded permuation operator between the
sites $k-1$ and $k+1$ with definition
$$\putequation{P}$$
Inspection of (\putlab{Hagain}) reveals the expressibility of the last
two terms in (\putlab{H4}) in terms of $u(1|2)$ generators
$J_{k,\ga}$, but the physical nature of the interctions is less
obvious in the resulting expression.\hfill\break
The results for $H_{(3)}$ and $H_{(4)}$ given in (\putlab{H3}) and
(\putlab{H4}) generalise trivially to the case of a lattice gas with
an $u(m|n)$ symmetry\upref{eks}/.

\vskip 1cm
\centerline{\bfs Acknowledgements:}
\hfill\break
We thank P.A. Bares, S. Dasmahapatra, B.A.S. Peeters, N.Yu.
Reshetikhin, E.K. Sklyanin and especially K. Schoutens for
discussions. This paper was typeset in \jyTeX .
\sectionnumstyle{Alphabetic}
\sectionnum=0
\equationnum=0
\begin{appendices}
{\bfs\section{Appendix : The Fundamental Representation of u$(1|2)$}}
In this appendix we write down the fundamental matrix representation
of $u(1|2)$ and use it to compute the inverse Killing form $K^{\ga\gb}$.
We consider the fundamental
matrix representation on site $k$, where we have chosen the fermionic
states to be
$\pmatrix{1&0&0\cr}^T$ (spin down electrons) and
$\pmatrix{0&1&0\cr}^T$ (spin up electrons) , and the bosonic state to
be $\pmatrix{0&0&1\cr}^T$ (empty site).
The generators are given by
$$\putequation{u12matrix}$$
The generators $J_{k,1},\ldots, J_{k,8}$ are seen to be
super-traceless, {\sl i.e.} $str(J) = J_{33}-J_{22}-J_{11} =0$.
The grading of the algebra is described in terms of Grassmann parities
$\epsilon _\alpha$, which are given by
$$\putequation{gp}$$
$K_{\ga\gb} = \left(K^{\ga\gb}\right)^{-1}$ is now given by the
following expression \upref Cornwell/
{\baselineskip=12pt $$\putequation{Kab}$$}
\vfill
{\bfs\section{Appendix : Computation of the ``unwanted terms'' for
the FFB grading}}
In this section we compute the so-called ``unwanted terms'' in the
expressions (\putlab{Dstate}), (\putlab{Astate}), (\putlab{ADstate1})
and (\putlab{ADstate2}).
The ``unwanted terms'' are characterised by containing a creation
operator $C$ with spectral parameter (SP) $\mu$ in place of a creation
operator with SP $\l{k}$ (or $\1l{k}$ for the nesting).
The condition of cancellation of these unwanted terms leads to the
Bethe equations.
In order to obtain the expression for ${\tilde \Lambda}_k$, we first
move the creation operator with SP $\l{k}$ to the first place in
(\putlab{state1}), using
(\putlab{AC})
$$\putequation{front}$$
To get an unwanted term, we now have to use the second term in
(\putlab{AC}) to move $D$ past $C_{b_k}(\l{k})$, and then always the
first term in (\putlab{AC}) to move $D$ (which now carries SP $\l{k}$)
to the very right, until it hits the vacuum, on which it acts according
to (\putlab{Tvac}).
This way we obtain
$$\putequation{laktilde}$$
The computation of $\Lambda_k$ is more complicated. We first derive an
expression for the contribution of $A_{11}(\mu )$, which we denote by
$\Lambda_{k,1}$. Proceeding along the same lines as in the computation
of ${\tilde \Lambda}_k$, we find
$$\putequation{lak1}$$
The $\gd_{d_{n-1},1}$ stems from the action of $A_{1,d_{n-1}}(\l{k})$
(which is what we get after moving $A$ past all the $C$'s) on the vacuum.
We also had to include a $\gd_{b_{k},1}$ due to the fact that in
(\putlab{Astate}) we denoted by $b_k$ the index of the $C$ with SP
$\mu$ (this means that on the l.h.s. of (\putlab{lak1}) $b_k$ is
really equal to $1$, too).
The contribution of $A_{22}(\mu )$ differs only in factors
$\gd_{b_{k},2}\gd_{d_{n-1},2}$ instead of
$\gd_{b_{k},1}\gd_{d_{n-1},1}$, so that the result for $\Lambda_k =
\Lambda_{k,1}+\Lambda_{k,2}$ is found to be
$$\putequation{lak2}$$
This expression can (and must be) simplified by carrying out the
contractions over the summation indices $c_1,\ldots ,c_k$.
Noting that
$$\putequation{invr}$$
we are able to perform all $c_i$-summations with the result
$$\putequation{Sr}$$
Now we transform the remaining $r$-matrices into $L$-operators, using
the identity
$$\putequation{raise}$$
The second equality holds, because $\ge_1=\ge_2=1$.
Thus we obtain our final form for the unwanted terms due to
$A_{11}(\mu )+A_{22}(\mu )$
$$\putequation{lak3}$$
We now insert (\putlab{lak3}) and (\putlab{laktilde}) into the
condition (\putlab{unw}) for the cancellation of the unwanted terms,
and multiply the resulting equation by the inverse of $S(\l{k})^{b_1\ldots
b_k}_{a_1\ldots a_k}$, which satisfies $\left(S^{-1}(\l{k})\right)^{p_1\ldots
p_k}_{b_1\ldots b_k}\ \times$ $\left(S(\l{k})\right)^{b_1\ldots
b_k}_{a_1\ldots a_k} = \prod_{i=1}^k \gd_{a_i,p_i}$, and which is
computed {\sl via}
(\putlab{invr}).
\vfill\hfill\break
After some trivial rearrangements we arrive at
$$\putequation{cancel}$$
This implies (\putlab{PBC}).\par
The unwanted terms (\putlab{ADstate1}) for the nesting can be computed
along similar lines and we easily find that
$$\putequation{1lak}$$
The cancellation of the unwanted terms for the nesting implies
(\putlab{pbc}).
{\bfs\section{Appendix : Equivalence of the BAE}}
In this appendix we establish the equivalence of our new set of BAE
(\putlab{BE32}) with the set obtained by Sutherland (\putlab{BE22}).
We use a method developed by Bares {\sl et.al.} in [\putref{Bares2}]
(see Appendix C of their paper) to show the equivalence of the
solutions of Sutherland and Lai (see also [\putref{Woynar}]).\par
We start by expressing the second set of our new BAE
$$\putequation{BAEnew}$$
as a polynomial equation of degree $N_h+N_\downarrow$
$$\putequation{pol}$$
Among the roots $w_j,\ j=1\ldots N_h+N_\downarrow$ of (\putlab{pol})
we consider $N_\downarrow$ roots $w_1,\ldots ,w_{N_\downarrow}$,
which we identify with $\1l{1},\ldots ,\1l{N_\da}$.
The $N_h$ other roots of (\putlab{pol}) we denote by ${w^\prime}_j$.
Using the residue theorem we can derive the following equality
$$\putequation{res}$$
where $C_j$ is a small contour around $w_j$.
The branch cut of the logarithm extends from $z_n=\tl{l}+{i\over 2}$
to $z_p=\tl{l}-{i\over 2}$. By deforming the contours on the r.h.s. of
(\putlab{res}) we arrive the the following equality
$$\putequation{res2}$$
where the last term on the r.h.s. comes from integration around the
branch cut. The form of the polynomial $p$ now implies that
$$\putequation{cut}$$
Inserting (\putlab{cut}) into (\putlab{res2}) and exponentiating the
result we obtain the identity
$$\putequation{id}$$
Now we use (\putlab{id}) in the first set of BAE in (\putlab{BAEnew})
with the result
$$\putequation{BAEsuth1}$$
This is precisely the first set of BAE in the Sutherland form
(\putlab{BE22}), if we make the identification ${w^\prime}_k =
\2t{k}$. The second set of the BAE (\putlab{BE22}) is also fullfilled
by the spectral parameters $\2t{k}$, because they are roots of the
equivalent polynomial equation (\putlab{pol}) (this is because the BAE
of the nesting for the Sutherland solution and the new solution are
identical).
Thus we have established the equivalence of the BAE (\putlab{BE32})
and (\putlab{BE22}).
\end{appendices}
\begin{putreferences}
\centerline{\bfs REFERENCES}
\vskip .5cm
\reference{Cornwell}{J.F. Cornwell, {\it Group Theory in Physics,
Vol III: Supersymmetries and Infinite-Dimensional Algebras}, Academic
Press (1989). The algebras $SU(n|m)$ are discussed on page 270.}
\reference{HaBe}{H.Bethe,\ \ZP{71}{1931}{205}.}
\reference{Yang}{C.N.Yang,\ \PRL{19}{1967}{1312}.}
\reference{Vlad}{V.E. Korepin, G. Izergin and N.M. Bogoliubov,
  {\it Quantum Inverse Scattering Method, Correlation Functions
  and Algebraic Bethe Ansatz}, Cambridge University Press, 1992 }
\reference{FT}{L.D.Faddeev, L. Takhtajan, Zap.Nauch.Semin LOMI Vol 109
(1981)\ p.134.}
\reference{FT2}{L.D.Faddeev in {\it Recent Advances in Field Theory
and Statistical Mechanics, Les Houches 1982}, page 561-608.}
\reference{Kul}{P.P.Kulish,\ \JSM{35}{1985}{2648}.}
\reference{KulRe}{P.P.Kulish, N.Yu. Reshetikhin,\ \JPA{16}{1983}{L591}.}
\reference{KulRe2}{P.P.Kulish, N.Yu. Reshetikhin,\ \JETP{53}{1981}{108}.}
\reference{BarBl}{P.A.Bares, G.Blatter,\ \PRL{64}{1990}{2567}.}
\reference{BarBlO}{P.A.Bares, G.Blatter, M.Ogata,\ \PRB{44}{1991}{130}.}
\reference{Bares}{P.A.Bares,\ PhD thesis, ETH Z\"urich (1991)}
\reference{Bares2}{P.A.Bares, J.M.P.Carmelo, J.Ferrer, P.Horsch,\ MIT
preprint, February 1992.}
\reference{Bill}{Bill Sutherland,\ \PRB{12}{1975}{3795}.}
\reference{Billy}{Bill Sutherland,\ \PRL{20}{1968}{98}.}
\reference{Lai}{C.K.Lai,\ \JMP{15}{1974}{1675}.}
\reference{Sarkar}{S.Sarkar,\ \JPA{24}{1991}{1137},\ \JPA{23}{1990}{L409}.}
\reference{Schlotti}{P.Schlottmann,\ \PRB{36}{1987}{5177}.}
\reference{Wiegmann}{P.B.Wiegmann,\ \PRL{60}{1988}{821}.}
\reference{Sklyanin}{E.K.Sklyanin,\ Univ. of Helsinki preprint
HU-TFT-91-51, Oct.1991}
\reference{Tetelman}{M.G.Tetel'man,\ \JETP{55}{1982}{306}.}
\reference{Taktajan}{L.Takhtajan,\ \JSM{23}{1983}{2470}.}
\reference{SklKul}{P.P.Kulish, E.K. Sklyanin,\ \JSM{19}{1982}{1596}.}
\reference{sup}{I.Bars, M.G\"unaydin,\ \CMP{91}{1983}{31}.}
\reference{Frahm}{H.Frahm,\ \JPA{25}{1992}{1417}.}
\reference{eks}{F.H.L.E\char'31 ler, V.E.Korepin, K.Schoutens,\
Stony Brook preprint ITP-SB 92-03.}
\reference{pw}{P.W.Anderson,\ \sci{235}{1987}{1196}.}
\reference{pw2}{P.W.Anderson,\ \PRL{65}{1990}{2306}.}
\reference{zhri}{F.C.Zhang, T.M.Rice,\ \PRB{37}{1988}{3759}.}
\reference{ek}{F.H.L.E\char'31 ler, V.E.Korepin,\ in preparation.}
\reference{Woynar}{F.Woynarovich,\ \JPC{16}{1983}{6593}.}
\end{putreferences}
\bye